\documentclass[aip,pof,twocolumn,longbibliography,reprint]{revtex4-1}

\usepackage{amssymb,amsmath}

\usepackage{epsfig}
\usepackage{graphicx}% Include figure files

\usepackage{color}

\newcommand\beq{\begin{equation}}
\newcommand\eeq{\end{equation}}
\newcommand\beqa{\begin{eqnarray}}
\newcommand\eeqa{\end{eqnarray}}
\newcommand{\al}{\alpha}

\usepackage{epstopdf}
\begin{document}
\title{Mpemba-like effect in driven binary mixtures}
\author{Rub\'en G\'omez Gonz\'alez}
\email{ruben@unex.es}
\affiliation{Departamento de
F\'{\i}sica, Universidad de Extremadura, E-06071 Badajoz, Spain}
\author{Nagi Khalil}
\email{nagi.khalil@urjc.es}
\affiliation{Escuela Superior de Ciencias Experimentales y Tecnolog\'ia (ESCET) \& GISC,
Universidad Rey Juan Carlos, M\'ostoles 28933, Madrid, Spain}
\author{Vicente Garz\'o}
\email{vicenteg@unex.es} \homepage{http://www.unex.es/eweb/fisteor/vicente/}
\affiliation{Departamento de F\'{\i}sica and Instituto de Computaci\'on Cient\'{\i}fica Avanzada (ICCAEx), Universidad de Extremadura, E-06071 Badajoz, Spain}

\begin{abstract}
The Mpemba effect occurs when two samples at different initial temperatures evolve in such a way that the temperatures cross each other during the relaxation towards equilibrium. In this paper we show the emergence of a Mpemba-like effect in a molecular binary mixture in contact with a thermal reservoir (bath). The interaction between the gaseous particles of the mixture and the thermal reservoir is modeled via a viscous drag force plus a stochastic Langevin-like term. The presence of the external bath couples the time evolution of the total and partial temperatures of each component allowing the appearance of the Mpemba phenomenon, even when the initial temperature differences are of the same order of the temperatures themselves. Analytical results are obtained by considering multitemperature Maxwellian approximations for the velocity distribution functions of each component. The theoretical analysis is carried out for initial states close to and far away (large Mpemba-like effect) from equilibrium. The former situation allows us to develop a simple theory where the time evolution equation for the temperature is linearized around its asymptotic equilibrium solution. This linear theory provides an expression for the crossover time. We also provide a qualitative description of the large Mpemba effect. Our theoretical results agree very well with computer simulations obtained by numerically solving the Enskog kinetic equation by means of the direct simulation Monte Carlo method and by performing molecular dynamics simulations. Finally, preliminary results for driven granular mixtures also show the occurrence of a Mpemba-like effect for inelastic collisions.
\end{abstract}

\draft

%\pacs{05.20.Dd, 45.70.Mg, 51.10.+y, 47.50.+d}
\date{\today}
\maketitle

%%%%%%%%%%%%%%%%%%%%%%%%%%%%%%%%%%%%%%%%%%%%%%%%
\section{Introduction}
\label{sec1}

The Mpemba effect is a counterintuitive phenomenon in which two samples of fluid $A$ and $B$ at initially different temperatures ($T_{A,0}>T_{B,0}$) can evolve in time in such a way that their temperatures cross each other at a given time $t_c$; the curve for $T_{A}$
(initially hotter sample) stays below the other one $T_{B}$ for longer times $t>t_c$. Although this anomalous cooling process was first reported in the case of water many years ago by E. B. Mpemba, \cite{Mpemba1969} its origin for that liquid is yet unclear. \cite{Kell1969,Deeson1971,Firth1971,Freeman1979,Auerbach1995,Maciejewski1996,Esposito2008,Katz2009,Vynnycky2010,Jin2015,Ibekwe2016,Mirabedin2017,
Keim2019,KRHV19,GLH19} In addition, although similar behaviors to the Mpemba effect have been observed in other systems,~\cite{Ahn2016,Hu2018,Kumar2020} the existence of the Mpemba effect still remains very controversial. \cite{Burridge2016,Burridge2020a} This is in part due to the arduous task of knowing and monitoring the initial conditions of the two samples that give rise to a crossing of the respective cooling curves.

For the above reason, in order to gain some insight into the understanding of this problem, the kinetic theory approach to granular gases \cite{Lasanta2017,Torrente2019,Biswas2020,THS21} has been widely employed in the past few years as a reliable tool for unveiling in a clean way the origin of the Mpemba-like effect (and its inverse one, namely, when initially cooler systems equilibrate faster than the hotter ones  \cite{Lu2017}) from a more fundamental point of view. Granular gases can be considered as a collection of macroscopic particles (typically of the orders of micrometers or larger) whose interactions are dissipative. The inelastic character of the collisions among granular particles gives rise to the coupling of the (granular) temperature with other velocity moments of the velocity distribution function, such as (i) the fourth cumulant or \emph{kurtosis} $a_2$ (a quantity measuring the departure of the distribution function from its Maxwellian form in driven granular gases),~ \cite{Lasanta2017} (ii) the rotational-to-translational temperature ratio in a granular gas of inelastic rough hard spheres, \cite{Torrente2019} (iii) the partial temperatures ratio in a binary granular mixture, \cite{Biswas2020} and (iv) the shear stress in a sheared inertial suspension.~\cite{THS21} The above couplings are the origin of the emergence of the Mpemba-like effect in granular gases, which now accounts for the evolution of the system towards a final asymptotic non-equilibrium steady state.

%%%%%%%%%%%%%%%%%%%%%%%%%%%%%%%%%%%%%%%%%%%%%%%%%%%%%%%%%%%%%%%%%%%%%%%%%%%%%%%%%%%%%%%%%%%%%%%%%%%%%%%%%%%%%%%%%%%%%%%%%%%%%%%%%%%%%%%%%%%%%

Among the above previous studies, to the best of our knowledge and in the context of kinetic theory, the only paper where the Mpemba-like effect has been studied for driven granular mixtures has been carried out by Biswas \emph{et al.}~\cite{Biswas2020} Since the number of parameters involved in multicomponent mixtures is much larger than that of a monocomponent gas, for the sake of simplicity, they consider the inelastic Maxwell model, namely, a simplified model for a granular gas where the collision rate is assumed to be independent of the relative velocity of the colliding particles. \cite{Garzo2019} The use of this simple model allows them to offer an exact analysis of the conditions under which the Mpemba-like effect is present. In addition, no simulations are performed in this work and only analytical results are reported. \cite{Biswas2020} Thus, given that inelastic Maxwell gases are an idealized version of the more realistic hard-sphere model, it is quite apparent that an study of the Mpemba-like effect in driven mixtures of hard spheres is still lacking.

%Here, we address this study by considering first molecular gases and then providing preliminary results for granular gases. In both cases, analytical approximate results are confronted against computer simulations showing an excellent agreement.

Aside granular gases, a recent paper \cite{Santos2020} has shown that the Mpemba effect can also take place in homogeneous and isotropic states of molecular gases (i.e. when collisions are elastic) in contact with a background fluid. The particles of the system are assumed to be hard spheres surrounded by an interstitial fluid at equilibrium. When the particles of the background fluid are much lighter than that of the gas (Brownian particles), the particles of the gas are subjected to a \emph{nonlinear} drag force plus a stochastic force with \emph{nonlinear} variance. After a transient period, the velocity distribution function is ensured to achieve a Maxwellian distribution with temperature given by the background fluid. To characterize the transient towards equilibrium, the first Sonine approximation to the distribution function (which includes the kurtosis $a_2$) was considered. As in the analysis performed in Ref.\ \onlinecite{Lasanta2017}, the Mpemba effect arises from the coupling of the time evolution of the temperature $T$ with that of the kurtosis $a_2$. This coupling is due here to the nonlinear form of the drag term. Nonetheless, as happens in the case of granular gases, \cite{Khalil2014,Garzo2019} since $a_2$ is small then the initial temperatures must be very close to each other in order to achieve a crossover in the evolution curves. As a consequence, the Mpemba crossover takes place in the very early stage of the relaxation towards equilibrium.

In this work, we analyze the occurrence of the Mpemba-like effect in a driven binary mixture of hard spheres. The driving of the mixture is due to its interaction with the surrounding molecular fluid. When the density of the gas is sufficiently low, one can assume that the interstitial fluid is not perturbed by the gas particles and so, it may be treated as a thermostat. As noted by Takada \emph{et al.}, \cite{THS21} the system we consider (inertial suspension) could be close to the original setup of Mpemba and Osborne~\cite{Mpemba1969} since they study a system of ice-mix, which is a suspension system.

Under the above conditions, as usual in granular literature, \cite{Koch2001} the interaction between gas particles and the surrounding fluid can be modeled by means of an effective external force. This fluid-solid interaction force (which follows the fluctuation--dissipation theorem \cite{Kawasaki2014,Hayakawa2017,THSG20}) is composed by two terms: (i) a \emph{linear} drag force proportional to the (instantaneous) velocity of particles  and (ii) a stochastic force. While the first contribution mimics the friction exerted by the viscous background fluid to the gas particles, the second term simulates the transmission of energy through random and instantaneous collisions with the external bath. This type of driving mechanism can be also formally obtained from the corresponding collision integral by considering the leading term in the Kramer--Moyal expansion in powers of the mass-ratio of the background and grain particles.~\cite{RL77,K92,zw01,BP04,OBB20}

In order to provide a simple explanation of the subjacent mechanisms involved in the Mpemba-like effect, we assume first a driven gas mixture of molecular gases (elastic collisions). This allow us to obtain simple conditions for the occurrence of such phenomenon. An extension of the study to inelastic collisions is briefly analyzed and illustrated in Sec.\ \ref{sec5}. However, given the complexity of the analytical expressions achieved for granular mixtures, it is not easy to provide simple conditions for the existence of the Mpemba-like effect for these systems.

As usual in driven mixtures, we consider the Enskog kinetic theory (which applies to small and moderate densities) in combination with the Fokker-Planck suspension model \cite{Koch1990} mentioned above. Starting from the set of Enskog kinetic equations for the mixture, the time evolution of the total temperature $T(t)$ and the partial temperatures $T_1(t)$ and $T_2(t)$ are obtained. As expected, the coupling between $T(t)$ and $T_1(t)$ and $T_2(t)$ is behind the emergence of the Mpemba-like memory effect. In addition, to get explicit results, the partial production rates $\xi_i$ (which give the rate of energy change in collisions $i$-$j$) appearing in the evolution equation of the partial temperatures are estimated here by assuming Maxwellian distributions at the temperatures $T_i$. This means that, in contrast to previous works,~\cite{Lasanta2017,Biswas2020,Santos2020} no cumulants nor the presence of a nonlinear drag force are needed for the emergence of the Mpemba effect.

Moreover, in accordance with simulations of bidisperse gas-solid flows,~\cite{Yin2009,Yin2009a,Holloway2009} the fact that the friction coefficients $\gamma_i$ accounting for the interaction of the component $i$ with the background fluid are different ($\gamma_1\neq \gamma_2$) makes different the energy transferred from the external bath to each component. As a consequence, the relaxation of $T_i$ towards its common equilibrium value ($T_1=T_2=T$) for molecular mixtures can be quite different for both partial temperatures. This makes the Mpemba effect arises even when the systems are initially prepared in Maxwellian velocity distribution functions at different partial temperatures. Moreover, the use of the partial temperatures as the control parameter allows some flexibility in the selection of the initial conditions and so, the magnitude of the Mpemba effect is not limited (namely, the so-called ``large'' Mpemba effect can be observed).

Nevertheless, given that the theoretical predictions derived here are based on a simple approximation (Maxwellian distributions for evaluating the production rates), a comparison with computer simulations turns out to be crucial to gauge their reliability. In this work, kinetic-theory results are compared against two independent simulation methods: (i) a modified algorithm of the standard Direct Monte Carlo Simulation (DSMC) method \cite{Bird1994}  to numerically solve the Enskog equation for a \emph{driven} binary mixture~\cite{Montanero1997,Montanero2002} and (ii) event-driven molecular dynamics (MD) simulations.~\cite{Khalil2014,allen2017computer,lu91} Both simulation methods complement each other since, on the one hand, DSMC offers a way to solve the Enskog equation by means of Monte-Carlo-like simulations. It inherently assumes the molecular chaos hypothesis and an approximate form of the pair distribution function at contact; both hypothesis stemming from the kinetic-theory description. At the same time, the DSMC method provides the \textit{exact} form of the time-dependent velocity distribution function, allowing us to assess the reliability of the approximate theory in the transient regime. On the other hand, for the suspension model considered here, given that MD solves numerically Newton's equations of motion with the action of a deterministic drag force plus a stochastic Langevin-like force, the limitations of the Enskog theory itself can be tested.

The paper is organized as follows. Section \ref{sec2} deals with the Enskog equation for homogeneous states conveniently adapted to the case of driven molecular binary mixtures. Evolution equations for the temperature ratio $T_1(t)/T_2(t)$ and the (total) temperature $T(t)$ are also derived. Next, Sec.\ \ref{sec3} analyzes states close to equilibrium. This allows us to linearize the above set of differential equations around the equilibrium solution and to solve them analytically. Exact expressions for the crossing time and the critical value of the initial temperature difference (which determines Mpemba and no Mpemba effects) are obtained and compared against DSMC simulations showing an excellent agreement. The large Mpemba-like effect is explored in Sec.\ \ref{sec4} in which we carry out a more qualitative analysis. Some examples regarding the fulfillment of the necessary but no sufficient conditions to achieve Mpemba effect are tested against both DSMC and MD simulations. Again, the theoretical results compare very well with computer simulations. Some preliminary results obtained for granular gases are presented in Sec.\ \ref{sec5} while the paper ends in Sec.\ \ref{sec6} with a brief discussion of the results derived in this work.

\section{Enskog kinetic theory for molecular binary mixtures in contact with a thermal reservoir}
\label{sec2}

Let us consider a binary mixture of hard particles of masses $m_1$ and $m_2$ and diameters $\sigma_1$ and $\sigma_2$. For the sake of simplicity and to provide a simple analysis of the conditions under which the Mpemba-like effect appears, we study first molecular binary mixtures (namely, when collisions between particles are elastic). Granular mixtures will be considered in Sec.\ \ref{sec5}. For moderate densities, the one-particle velocity distribution function $f_i(\mathbf{r}, \mathbf{v},t)$ ($i=1,2$) of the component $i$ obeys the Enskog kinetic equation.~\cite{SydneyChapman1991} We assume that the mixture interacts with a thermal reservoir (or equivalently, particles of the gas are surrounded by an interstitial fluid) so that the total temperature of the mixture does not remain constant and changes in time. To model the interaction of the particles of the gas with the surrounding fluid, one possibility would be to describe the molecular suspension in terms of a set of two coupled kinetic equations for each one of the velocity distributions of the different phases. However, the resulting theory would be very difficult to solve, specially in the case of multicomponent mixtures. For this reason, due to the technical difficulties involved in the above approach, it is more usual in gas-solid flows to model the influence of the interstitial fluid on particles of the gas mixture by means of an effective external force.~\cite{Koch2001}

For \emph{homogeneous} and isotropic states, the set of Enskog coupled equations reads~\cite{SydneyChapman1991,Cercignani1988}
\beq
\label{1}
\frac{\partial f_i}{\partial t}=\sum_{j=1}^2J_{ij}[\mathbf{v}|f_i,f_j]+\mathcal{C}_{i,\text{ex}}, \quad (i=1,2)
\eeq
where the Boltzmann--Enskog collision operator $J_{ij}[f_i,f_j]$ for homogeneous states is \cite{SydneyChapman1991,Cercignani1988}
\beqa
\label{1.1}
J_{ij}[\mathbf{v}_1|f_i,f_j]&=&\chi_{ij} \sigma_{ij}^{d-1}\int d\mathbf{v}_{2}\int d\widehat{\boldsymbol {\sigma
}}\,\Theta (\widehat{{\boldsymbol {\sigma }}} \cdot {\mathbf g}_{12}) (\widehat{\boldsymbol {\sigma }}\cdot {\mathbf g}_{12}) \nonumber \\
& & \times
\left[f_{i}({\mathbf v}_{1}')f_{j}({\mathbf v}_{2}')-f_{i}({\mathbf v}
_{1})f_{j}({\mathbf v}_{2})\right].
\eeqa
Here, $\sigma_{ij}=\left( \sigma_{i}+\sigma_{j}\right) /2$,
$\widehat{\boldsymbol {\sigma}}$ is an unit vector along the line of centers, $\Theta $ is the Heaviside step function, $d$ is the dimensionality of the system ($d = 2$ for disks and $d = 3$ for spheres),
$\mathbf{g}_{12}={\mathbf v}_{1}-{\mathbf v}_{2}$ is the relative velocity, and the relation between the pre-collisional velocities $(\mathbf{v}_1,\mathbf{v}_2)$ and the post-collisional velocities $(\mathbf{v}_1',\mathbf{v}_2')$ is
\beq
\label{1.1.0}
\mathbf{v}_{1}'=\mathbf{v}_{1}-2\mu_{ji}(\widehat{{\boldsymbol {\sigma }}} \cdot {\mathbf g}_{12})\widehat{\boldsymbol {\sigma}}, \quad
\mathbf{v}_{2}'=\mathbf{v}_{1}+2\mu_{ij}(\widehat{{\boldsymbol {\sigma }}} \cdot {\mathbf g}_{12})\widehat{\boldsymbol {\sigma}},
\eeq
where $\mu_{ij}=m_i/(m_i+m_j)$.  Moreover, $\chi_{ij}$ is the pair correlation function at thermal equilibrium for particles of types $i$ and $j$ when they are in contact, i.e. separated by $\sigma_{ij}$. Except for the presence of the pair correlation functions $\chi_{ij}$, the Enskog equation for homogeneous states is identical to the Boltzmann equation.

The second term $\mathcal{C}_{i,\text{ex}}$ of the right-hand side of Eq.\ \eqref{1} describes the coupling between the thermal reservoir and particles of the component $i$. As said in the Introduction, if the gaseous mixture is sufficiently dilute, one can neglect the impact of gas particles on the surrounding fluid and so, the latter plays the role of a thermostat. In this case, a reliable model for describing suspensions is the Langevin equation, ~\cite{Hayakawa2017,THSG20,THS21} so that the influence of the background fluid on gas particles is accounted for (i) a deterministic viscous (linear) drag force proportional to the particle velocity \cite{Koch2001} plus (ii) a stochastic Langevin force representing Gaussian white noise. \cite{Kampen1981} This latter term is represented by a Fokker--Planck collision operator. \cite{Noije1998,Henrique2000,Dahl2002} While the drag force term models the friction of particles of the component $i$ with the surrounding fluid, the stochastic term attempts to mimic the energy gained by particles of the gas due to their interactions with the more rapid particles of the interstitial fluid. Thus, the term $\mathcal{C}_{i,\text{ex}}$ reads
\beq
\label{1.1.1}
\mathcal{C}_{i,\text{ex}}=\gamma_i\frac{\partial}{\partial\mathbf{v}}\cdot\mathbf{v}f_i+\frac{\gamma_i T_{\text{ex}}}{m_i}\frac{\partial^2 f_i}{\partial v^2},
\eeq
where the coefficients $\gamma_i$ are the friction or drift coefficients and $T_{\text{ex}}$ is the background temperature. Here, we have taken units for the temperature for which the Boltzmann constant $k_\text{B}=1$. The structure of Eq.\ \eqref{1.1.1} can be also derived from the Boltzmann--Lorentz collision operator (characterizing the effect of collisions on the distribution $f_i$ between the Brownian particle $i$ and fluid particles) by considering the leading term in the Kramers--Moyal expansion in powers of the mass ratio $m_f/m_i$ when the background fluid is at equilibrium.~\cite{RL77,K92,zw01,BP04} Here, $m_f$ denotes the mass of the particles of the background fluid.

It must be noted that in general the friction coefficients may be tensorial quantities as a result of the hydrodynamic interactions between particles, which strongly depends on the configuration of particles. Here, the isotropic case is considered for the sake of simplicity and so, the coefficients $\gamma_i$ are scalar quantities. Thus, in the case of granular particles, the suspension model employed here might be applicable to describe inertial suspensions where the diameter of suspended particles ranges approximately from 1 to 70 $\mu$m.~\cite{Koch2001}

Under the above conditions, the Enskog--Fokker--Planck kinetic equation \eqref{1} is~\cite{Gonzalez2020}
\beq
\label{2}
\frac{\partial f_i}{\partial t}-\gamma_i\frac{\partial}{\partial\mathbf{v}}\cdot\mathbf{v}f_i-\frac{\gamma_i T_{\text{ex}}}{m_i}\frac{\partial^2 f_i}{\partial v^2}=\sum_{j=1}^2\; J_{ij}[f_i,f_j].
\eeq
The coefficients $\gamma_i$ can be written as $\gamma_i=\gamma_0 R_i$, where $\gamma_0\propto \sqrt{T_\text{ex}}$. The dimensionless quantities $R_i$ may depend on the mass ratio $m_1/m_2$, the diameter ratio $\sigma_1/\sigma_2$, the total volume fraction $\phi=\phi_1+\phi_2$, and the partial volume fractions $\phi_i$ defined as
\beq
\label{2.1}
\phi_i=\frac{\pi^{d/2}}{2^{d-1}d\Gamma\Big(\frac{d}{2}\Big)}n_i\sigma_i^{d}.
\eeq
The suspension model \eqref{2} has been recently  employed to determine the Navier--Stokes transport coefficients of bidisperse granular suspensions~\cite{Gonzalez2020} as well as the rheological properties in inertial suspensions of inelastic rough hard spheres under simple shear flow.~\cite{GGG20}

Explicit forms of $R_i$ have been displayed in the literature of polydisperse gas-solid flows.~\cite{Yin2009,Yin2009a,Holloway2009} In particular, we adopt the expression $\gamma_i=(18 \eta_g/\rho \sigma_{12}^2)R_i$ proposed by Yin and Sundaresan.~\cite{Yin2009a} Here, $\rho=\sum_i m_i n_i$ is the total mass density and
\beq
\label{4}
n_i=\int d \mathbf{v}\; f_i(\mathbf{v})
\eeq
is the number density of the component $i$. For a three-dimensional low-Reynolds-number fluid at moderate densities, the dimensionless function $R_i$ is given by
\beqa
\label{2.1.1}
R_i(\phi_i,\phi)&=&\frac{\rho \sigma_{12}^2}{\rho_i \sigma_i^2}\frac{(1-\phi)\phi_i\sigma_i}{\phi}\sum_{j=1}^{2}\frac{\phi_j}{\sigma_j}\Bigg[
\frac{10\phi}{\left(1-\phi\right)^2}
\nonumber\\
& & +\left(1-\phi\right)^2\left(1+1.5\sqrt{\phi}\right)\Bigg].
\eeqa

At a kinetic level, one of the most relevant quantities for a binary mixture are the partial temperatures $T_i(t)$. They measure the mean kinetic energy of the component $i$ and are defined as
\beq
\label{3}
T_i=\frac{1}{n_i d}\int d \mathbf{v}\; m_i v^2\; f_i(\mathbf{v}).
\eeq
Alternatively, the same information is provided by the temperature ratio $\theta(t)=T_1(t)/T_2(t)$ and the (total) temperature $T(t)$ of the mixture
\beq
\label{5}
T(t)=x_1 T_1(t)+x_2 T_2(t),
\eeq
where $x_i=n_i/(n_1+n_2)$ is the mole fraction of the component $i$. The ratios $T_1/T$ and $T_2/T$ can be easily expressed in terms of $\theta$ as
\beq
\label{6}
\frac{T_1}{T}=\frac{\theta}{1+x_1(\theta-1)}, \quad \frac{T_2}{T}=\frac{1}{1+x_1(\theta-1)}.
\eeq

The evolution equations for both the temperature ratio $\theta$ and the (total) temperature $T$ can be obtained by multiplying both sides of the Enskog equation \eqref{2} by $m_i v^2$ and integrating over velocity. They are given by
\beq
\label{6b}
\frac{\partial}{\partial t}\ln T= 2 x_1 \gamma_1 \frac{T_\text{ex}-T_1}{T}+2 x_2 \gamma_2 \frac{T_\text{ex}-T_2}{T},
\eeq
\beq
\label{7}
\frac{\partial}{\partial t}\ln \theta=2\gamma_1\left(\frac{T_\text{ex}}{T_1}-1\right)-2\gamma_2\left(\frac{T_\text{ex}}{T_1}\theta-1\right)
+\xi_2-\xi_1,
\eeq
where
\beq
\label{8}
\xi_i=-\frac{m_i}{d n_i T_i}\int d \mathbf{v}\; v^2 J_{ij}[f_i,f_j], \quad (i\neq j),
\eeq
are the so-called partial production rates. They measure the rate of change of the kinetic energy of the particles of component $i$ due to collisions with the particles of component $j$. Since the collisions are elastic, we have $x_1 T_1 \xi_1+x_2 T_2 \xi_2=0$.

In the particular case of mechanically equivalent particles ($m_1=m_2$, $\sigma_1=\sigma_2$, and $\phi_1=\phi_2$), the friction coefficients $\gamma_1=\gamma_2=\gamma$ and the solution to Eq.\ \eqref{7} is simply
\beq
\label{8.1}
T(t)=T_\text{ex}+\left[T(0)-T_\text{ex}\right]e^{-2\gamma t}.
\eeq
Thus, since $\gamma>0$, the temperature decays monotonically in time and the Mpemba effect is not present.  However, when both components are different ($\gamma_1\neq \gamma_2$), the evolution equations of $T(t)$ and $\theta(t)$ are coupled: the curve of the initially hotter (cooler) sample  may cross that of the initally cooler (hotter) one and remain below (above) it until the systems reach equilibrium. This is the usual (or \emph{inverse}) Mpemba-like effect.

According to Eq.\ \eqref{8}, it is quite apparent that one needs to know the velocity distributions $f_1$ and $f_2$ to determine the partial production rates $\xi_1$ and $\xi_2$. Here, we estimate both production rates by taking the simplest approximation for the distributions $f_1$ and $f_2$, namely the Maxwellian distributions $f_{i,\text{M}}$ defined with partial temperatures $T_i$:
\beq
\label{9}
f_{i,\text{M}} (\mathbf{v};t)=n_i \left(\frac{m_i}{2\pi T_i(t)}\right)^{d/2} \exp\left(-\frac{m_i v^2}{2 T_i(t)}\right).
\eeq
In the Maxwellian approximation, $\xi_1$ is given by~\cite{Holway1966,Goldman1967,garzo2003kinetic}
\beqa
\label{10}
\xi_{1}&=&\frac{8\pi^{\left(d-1\right)/2}}{d\Gamma\left(\frac{d}{2}\right)}n_2 \mu_{12}\mu_{21}\chi_{12} \sigma_{12}^{d-1}\nonumber \\ & &
\times \left(\frac{2 T_1}{m_1}+\frac{2 T_2}{m_2}
\right)^{1/2}\left(1-\frac{T_2}{T_1}\right).
\eeqa
The expression of $\xi_2$ can be easily inferred from Eq.\ \eqref{10} by making the change $1\leftrightarrow 2$ in the sub-indexes. When energy equipartition holds ($T_1=T_2$), $\xi_1=\xi_2=0$ as expected. However, if energy equipartition is broken ($T_1\neq T_2$), then $\xi_i\neq 0$.

In the long-time limit, the mixture achieves a equilibrium state where energy equipartition applies: $T^\text{eq}_{1}=T^\text{eq}_{2}=T_\text{eq}=T_{\text{ex}}$. However, in the transient regime, it is expected that energy equipartition fails and so, $T_1(t)\neq T_2(t)$. This means that the Mpemba effect in a driven molecular mixture stems from the non-equipartition of energy. Remarkably, this effect can be explained by computing $\xi_i$ by a Maxwellian distribution and hence, the existence of different partial temperatures is sufficient to explain such memory effect.

In order to analyze the time dependence of $T(t)$ and $\theta(t)$ it is convenient to introduce dimensionless variables for temperature and time. Thus, we define $T^*=T/T_\text{ex}$ and $t^*=n\sigma_{12}^{d-1}\sqrt{4 T_\text{ex}/(m_1+m_2)} t$. In the Maxwellian approximation, the evolution equations for $T^*$ and $\theta$ can be easily derived from Eqs.\ \eqref{6b}, \eqref{7}, and \eqref{10}. After some algebra, one gets
\beq
\label{11}
\frac{\partial}{\partial t^*}\ln T^*=\Phi(T^*,\theta), \quad \frac{\partial}{\partial t^*}\ln \theta=\Psi(T^*,\theta),
\eeq
where
\beqa
\label{12}
\Phi(T^*,\theta)&=&\Phi_1(T^*)+\Phi_2(\theta),
\nonumber  \\ \Psi(T^*,\theta)&=&\Psi_1+\Psi_2(T^*,\theta)+\Psi_3(T^*,\theta).
\eeqa
Here, we have introduced the quantities
\begin{widetext}
\beq
\label{13}
\Phi_1(T^*)=\frac{2}{T^*}\left(x_1\gamma_1^*+x_2 \gamma_2^*\right), \quad \Phi_2(\theta)=-2 \frac{x_1 \gamma_1^* \theta+x_2 \gamma_2^*}{1+x_1(\theta-1)},
\eeq
\beq
\label{14}
\Psi_1=-2(\gamma_1^*-\gamma_2^*), \quad \Psi_2(T^*,\theta)=2\left(\gamma^*_1-\gamma_2^*\theta\right)\frac{1+x_1(\theta-1)}{\theta T^*},
\eeq
\beq
\label{15}
\Psi_3(T^*,\theta)=\frac{8\pi^{\left(d-1\right)/2}}{d\Gamma\left(\frac{d}{2}\right)}\chi_{12}\sqrt{\frac{T^*}{2}
\frac{\mu_{12}\mu_{21}\left(\mu_{12}+\mu_{21}\theta\right)}{1+x_1(\theta-1)}}
\left(x_1-x_2-x_1\theta+x_2 \theta^{-1}\right),
\eeq
where
\beq
\label{15.1}
\gamma_i^*=\frac{R_i}{\sqrt{2T_\text{ex}^*}(n_1+n_2)\sigma_{12}^d}, \quad T_\text{ex}^*=\frac{2T_\text{ex}}{(m_1+m_2)\sigma_{12}^2\gamma_0^2}.
\eeq
As mentioned before, the dependence of $\Phi$ on $\theta$ is a necessary condition for the existence of the Mpemba-like effect.
\end{widetext}

%%%%%%%%%%%%%%%%%%%%%%%%%%%%%%%%%%%%%%%%%%%%%%%%%%%%%%%%%%%%%%%%%%%%%%%%%

\section{Mpemba-like effect for initial states close to equilibrium}
\label{sec3}

We consider two homogeneous states A and B characterized by their initial reduced temperatures $T^*_{I,0}$ and temperature ratios $\theta_{I,0}$, where $I=\text{A,B}$. For the sake of simplicity, we suppose that both states are hotter (cooler) than the equilibrium state, i.e. $T_{A,0}^*>1$ and $T_{B,0}^*>1$ ($T_{A,0}^*<1$ and $T_{B,0}^*<1$). Furthermore, we also assume that $T_{\text{A},0}^*>T_{\text{B},0}^*>1$ ($T_{\text{A},0}^*<T_{\text{B},0}^*<1$ for the cooler case). During the time evolution of the system towards equilibrium the gas particles exchange energy with the thermal reservoir. This interaction is controlled by the friction coefficients $\gamma_i\propto R_i$, which exhibit a complex dependence on the mass and diameter ratios and the composition [see Eq.\ \eqref{2.1.1}]. Thus, the energy transfer (per particle) between each one of the components of the mixture and the background fluid could be more efficient (larger) for some values of $m_i$, $\sigma_i$, and $x_i$. So, as the energy transmission distinguishes between both components, the decay of the temperature until its equilibrium value will depend separately on the way of releasing energy from each component of the mixture to the bath, and consequently on the initial values of the partial temperatures $\theta_{I,0}$. This coupling between $T^*$ and $\theta$ opens up the possibility of a crossroad between the trajectories of both temperatures (Mpemba-like effect), so that $T_\text{A}^*=T_\text{B}^*$ at some crossing time $t_c^*$ before achieving the equilibrium state.

In order to quantify the constraints in the initial conditions of both trajectories needed for
the existence of $t_c^*$, we consider first in this Section initial states that are close to the final equilibrium state. Under these conditions, Eqs. \eqref{11} can be linearized around the equilibrium solution $T_\text{eq}^*=\theta_\text{eq}=1$. Note that this is a special kind of linearization since \emph{only} the global temperature and the temperature ratio are displaced with respect to their equilibrium values. As we show later, this approach will allow us to solve the corresponding set of linear differential equations and get analytical results.
	
Let us define  $\delta T^*=T^*-1$ and $\delta\theta=\theta-1$. Substitution of these definitions into Eqs.\ \eqref{11} and retaining only linear terms in $\delta T^*$ and $\delta\theta$, one gets
\beq
\label{16}
\frac{\partial}{\partial t^*} \begin{pmatrix} \delta T^* \\ \delta \theta \end{pmatrix}=\mathcal{L}\begin{pmatrix} \delta T^* \\ \delta \theta \end{pmatrix},
\eeq
where the matrix $\mathcal{L}$ is composed by the following elements:
\beqa
\label{17}
\mathcal L_{11}&=&-2(x_1 \gamma_1^*+x_2 \gamma_2^*), \nonumber \\
\mathcal L_{12}&=&2x_1x_2(\gamma_2^*-\gamma_1^*),\quad \mathcal L_{21}=2(\gamma_2^*-\gamma_1^*),\quad \nonumber\\ \mathcal L_{22}&=&-2(x_2\gamma_1^*+x_1\gamma_2^*)-\frac{4\pi^{(d-1)/2}}{d\Gamma\left(\frac{d}{2}\right)}\chi_{12}
\sqrt{2\mu_{21}\mu_{12}}.\nonumber\\
\eeqa

The solution to the matrix equation \eqref{16} can be expressed in terms of the initial conditions $\delta T^*_0$ and $\delta\theta_0$. After some algebra, the time evolution of the temperature $\delta T^*(t^*)$ reads
\beqa
\label{18}
\delta T^*(t^*)&=&\frac{1}{\lambda_{+}-\lambda_{-}}\bigg\{\left[\left(\mathcal L_{11}-\lambda_{-}\right)\delta T^*_0+\mathcal L_{12}\delta\theta_0\right]\operatorname{e}^{\lambda_{+}t^*}
\nonumber \\ & & +\left[\left(\lambda_{+}-\mathcal L_{11}\right)\delta T^*_0-\mathcal L_{12}\delta\theta_0\right]\operatorname{e}^{\lambda_{-}t^*}\bigg\},
\eeqa
where
\beq
\label{18.1}
\lambda_{\pm}=\frac{1}{2}\left[\mathcal L_{11}+\mathcal L_{22}\pm\sqrt{\left(\mathcal L_{11}-\mathcal L_{22}\right)^2+4\mathcal L_{12}\mathcal L_{21}}\right]
\eeq
are the eigenvalues of the matrix $\mathcal{L}$. Two observations are in order here. On the one hand, $\lambda_\pm\le 0$ for any choice of the system parameters since $\mathcal L_{11}+\mathcal L_{22}\le 0$ and
\beqa
\label{18.2}
& & \left(\mathcal L_{11}+\mathcal L_{22}\right)^2 -\Big[\left(\mathcal L_{11}-\mathcal L_{22}\right)^2+4\mathcal L_{12}\mathcal L_{21}\Big]=
8\Big[2\gamma_1^*\gamma_2^*\nonumber\\
& &+\frac{4\pi^{(d-1)/2}}{d\Gamma\left(\frac{d}{2}\right)}\chi_{12}
\sqrt{2\mu_{21}\mu_{12}}\left(x_1\gamma_1^*+x_2\gamma_2^*\right)\Big]\ge 0.
\eeqa
Hence, Eq.~\eqref{18} always describes an evolution of the system towards thermal equilibrium. On the other hand, from Eq.~\eqref{18} it is obvious that any cooling process ($\delta T^*\ge 0$) has its associated heating process ($\delta T^*\le 0$). One process can be obtained from the other one by a change of sings on the initial conditions: $\delta T^*_0\leftrightarrow -\delta T^*_0$ and $\delta\theta_0\leftrightarrow -\delta\theta_0$. Hence, at the level of the linear theory, the Mpemba effect occurs if and only if the inverse Mpemba effect occurs, provided the initial conditions are related by the previous sign transformation.

\begin{figure}[h]
	\centering
	\includegraphics[width=0.45\textwidth]{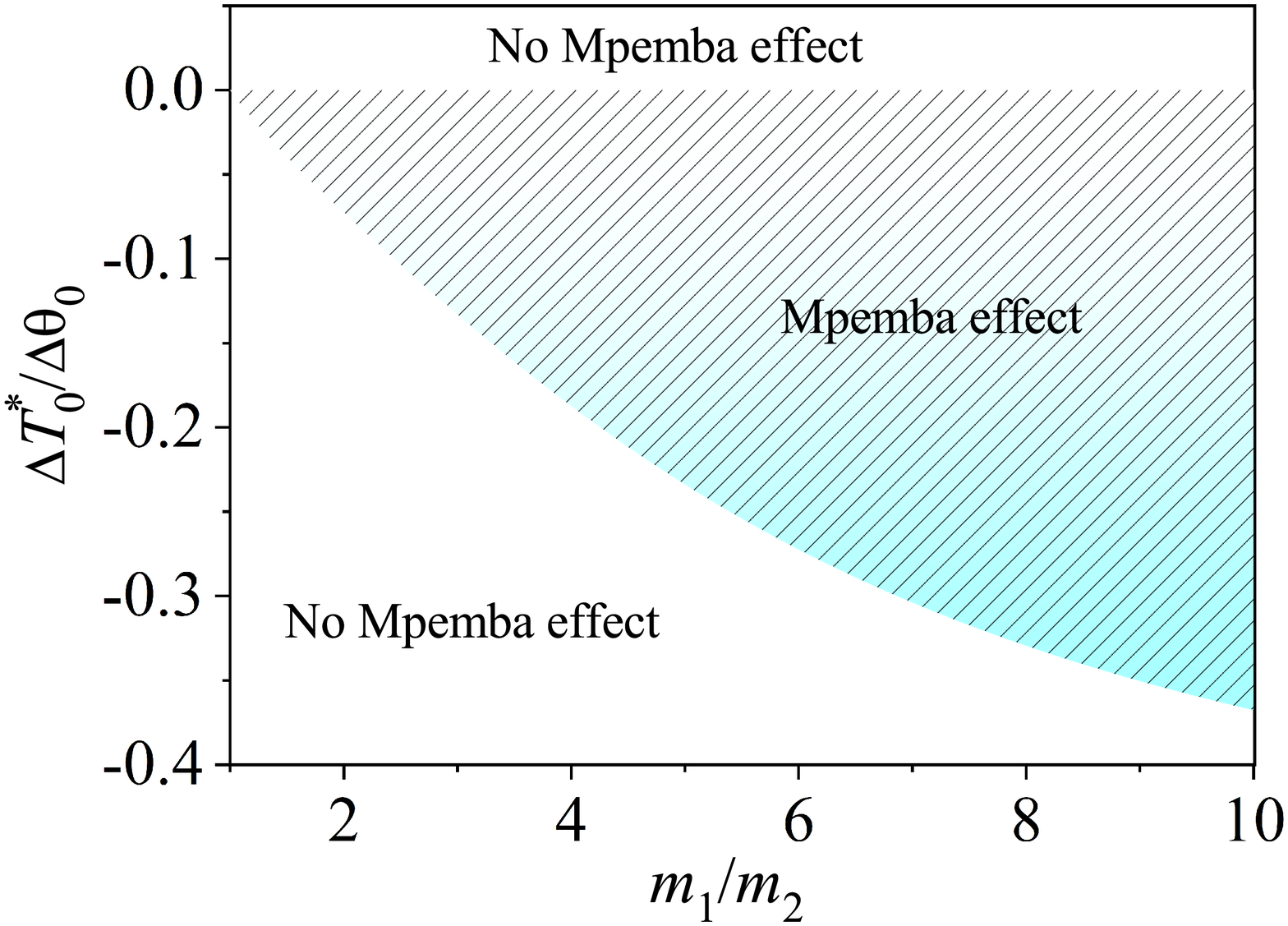}
	\includegraphics[width=0.45\textwidth]{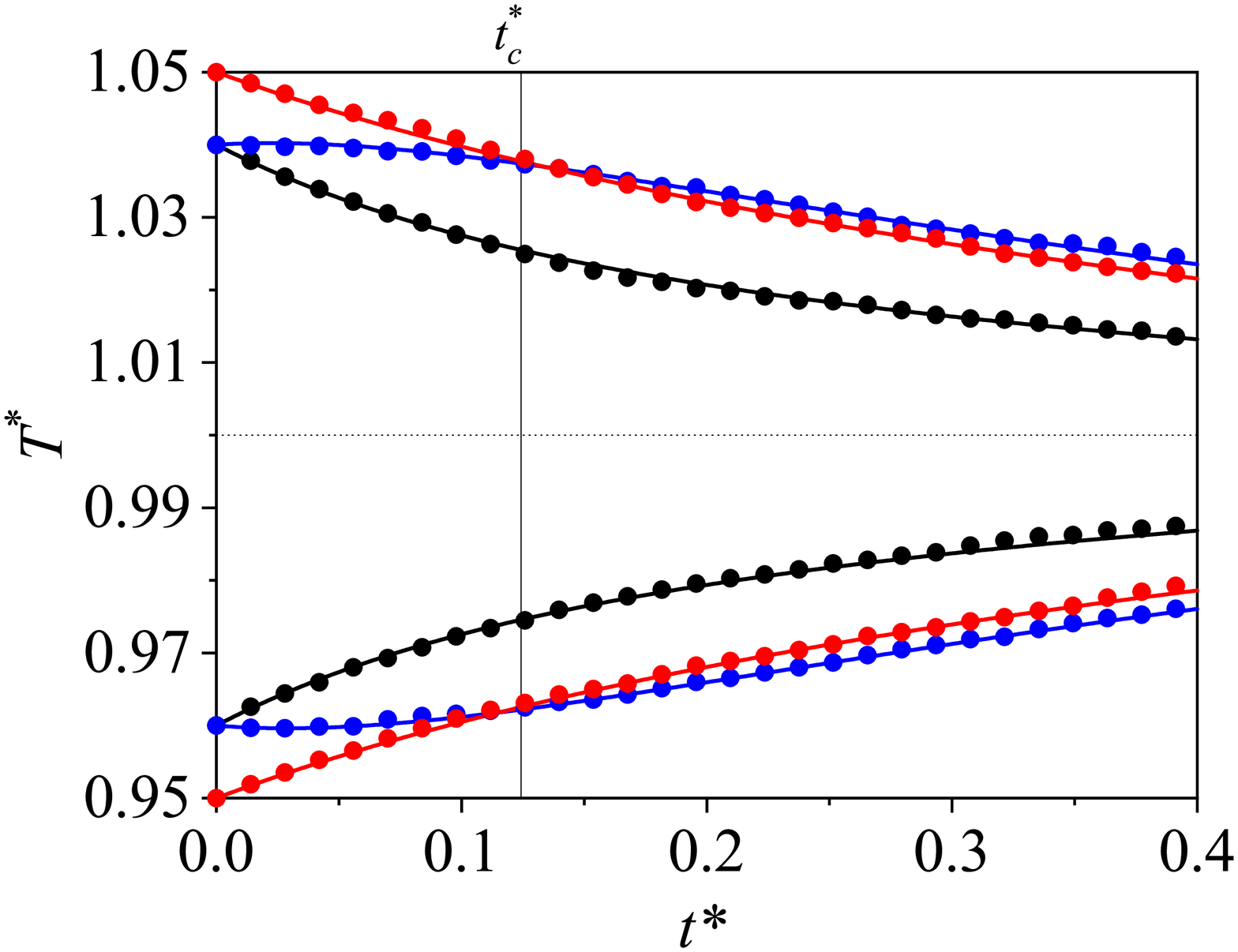}
	\caption{Upper panel: Phase diagram of the initial condition $\Delta T^*_0/\Delta\theta_0$ as a function of the mass ratio $m_1/m_2$. Lower panel: Relaxation of the (reduced) temperature $T^*$ over the time $t^*$ for $m_1/m_2=10$. The upper and lower curves correspond to the cooling and heating cases, respectively. Solid lines represents theoretical values and symbols DSMC data. The initial values of the control parameter $\Delta T^*_0/\Delta\theta_0\equiv (T^*_{A,0}-T^*_{B,0})/(\theta_{A,0}-\theta_{B,0})$ are $0.2$ (A: red line and symbols; B: black lines and symbols), and $-0.2$ (A: red lines and symbols; B: blue lines and symbols). The theoretical value of $t^*_c$ is also plotted with a vertical line. The remaining parameters in both panels are $d=3$,  $T_\text{ex}^*=1$, $x_1=\frac{1}{2}$, $\sigma_1/\sigma_2=1$, and $\phi=0.1$.}
	\label{fig1}
\end{figure}

\begin{figure}[h]
	\centering
	\includegraphics[width=0.45\textwidth]{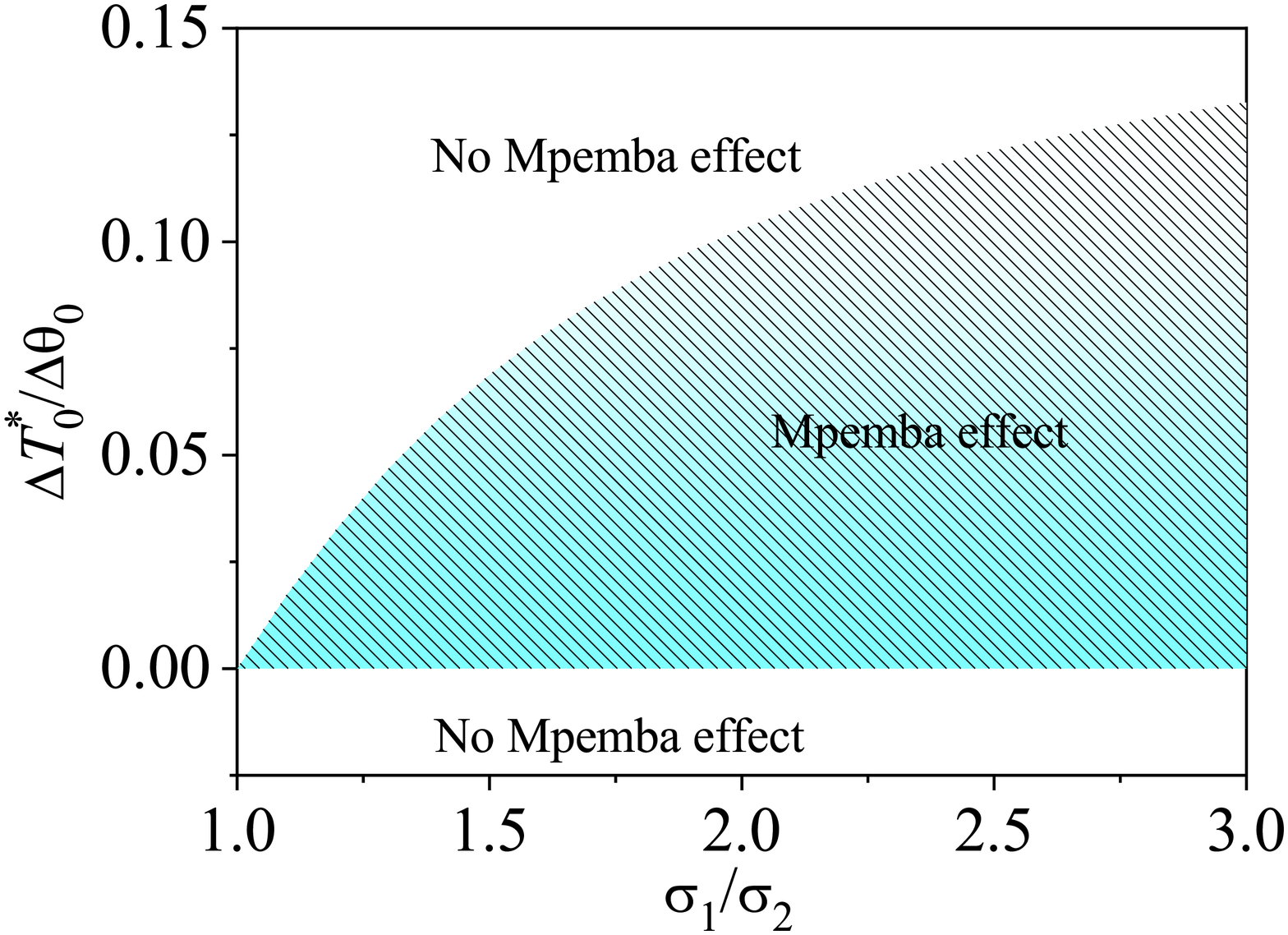}
	\includegraphics[width=0.45   \textwidth]{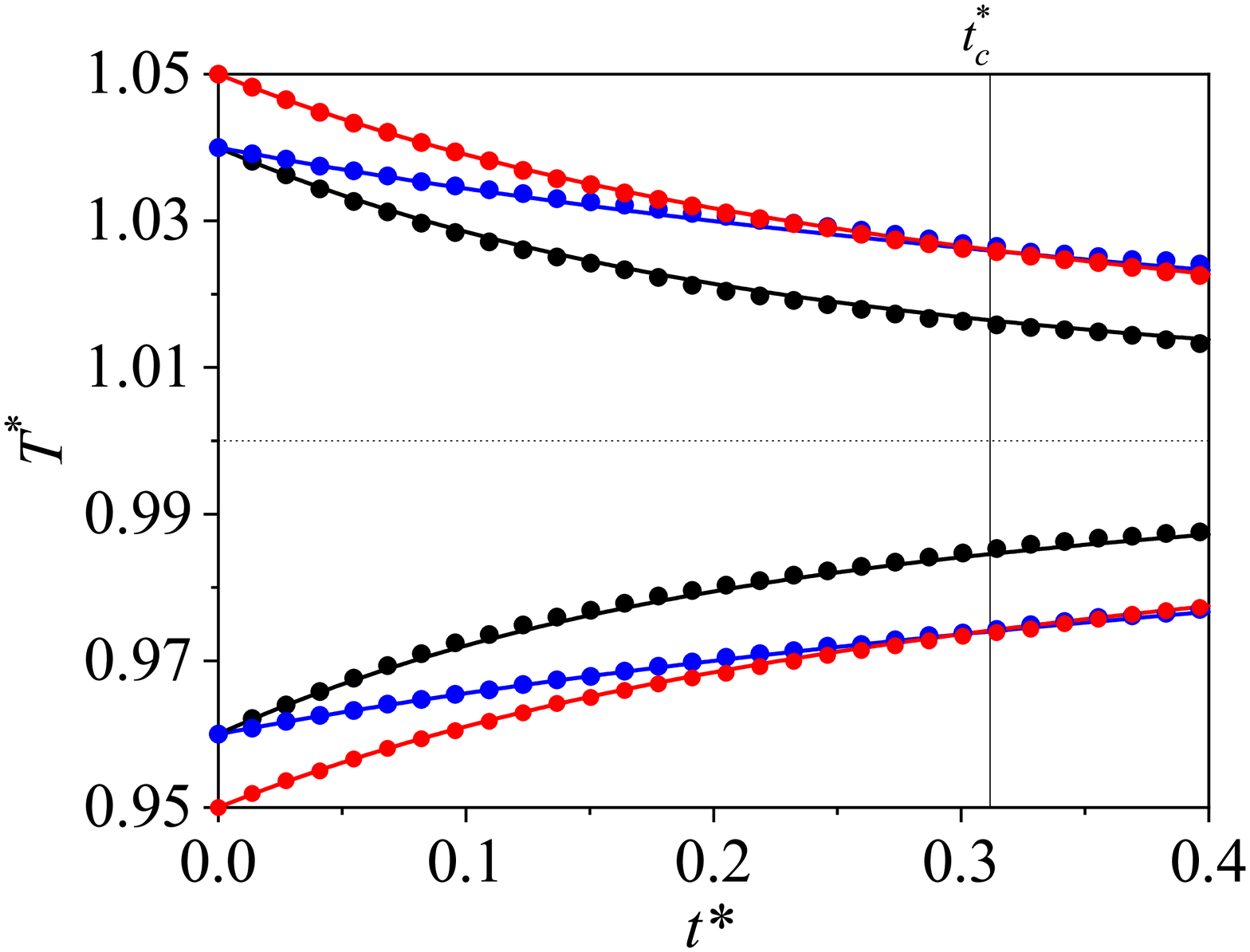}
	\caption{Upper panel: Phase diagram of the initial condition $\Delta T^*_0/\Delta\theta_0$ as a function of the size ratio $\sigma_1/\sigma_2$. Lower panel: Relaxation of the (reduced) temperature $T^*$ over the time $t^*$ for $\sigma_1/\sigma_2=3$. The upper and lower curves correspond to the cooling and heating cases, respectively. Solid lines represents theoretical values and symbols DSMC data. The initial values of the control parameter $\Delta T^*_0/\Delta\theta_0\equiv (T^*_{A,0}-T^*_{B,0})/(\theta_{A,0}-\theta_{B,0})$ are $0.1$ (A: red lines and symbols; B: blue lines and symbols), and $-0.2$ (A: red lines and symbols; B: black lines and symbols). The theoretical value of $t^*_c$ is also plotted with a vertical line. The remaining parameters in both panels are $d=3$, $T_\text{ex}^*=1$, $x_1=\frac{1}{2}$, $m_1/m_2=1$, and $\phi=0.1$.}
	\label{fig2}
\end{figure}

\begin{figure}[h]
	\centering
	\includegraphics[width=0.45\textwidth]{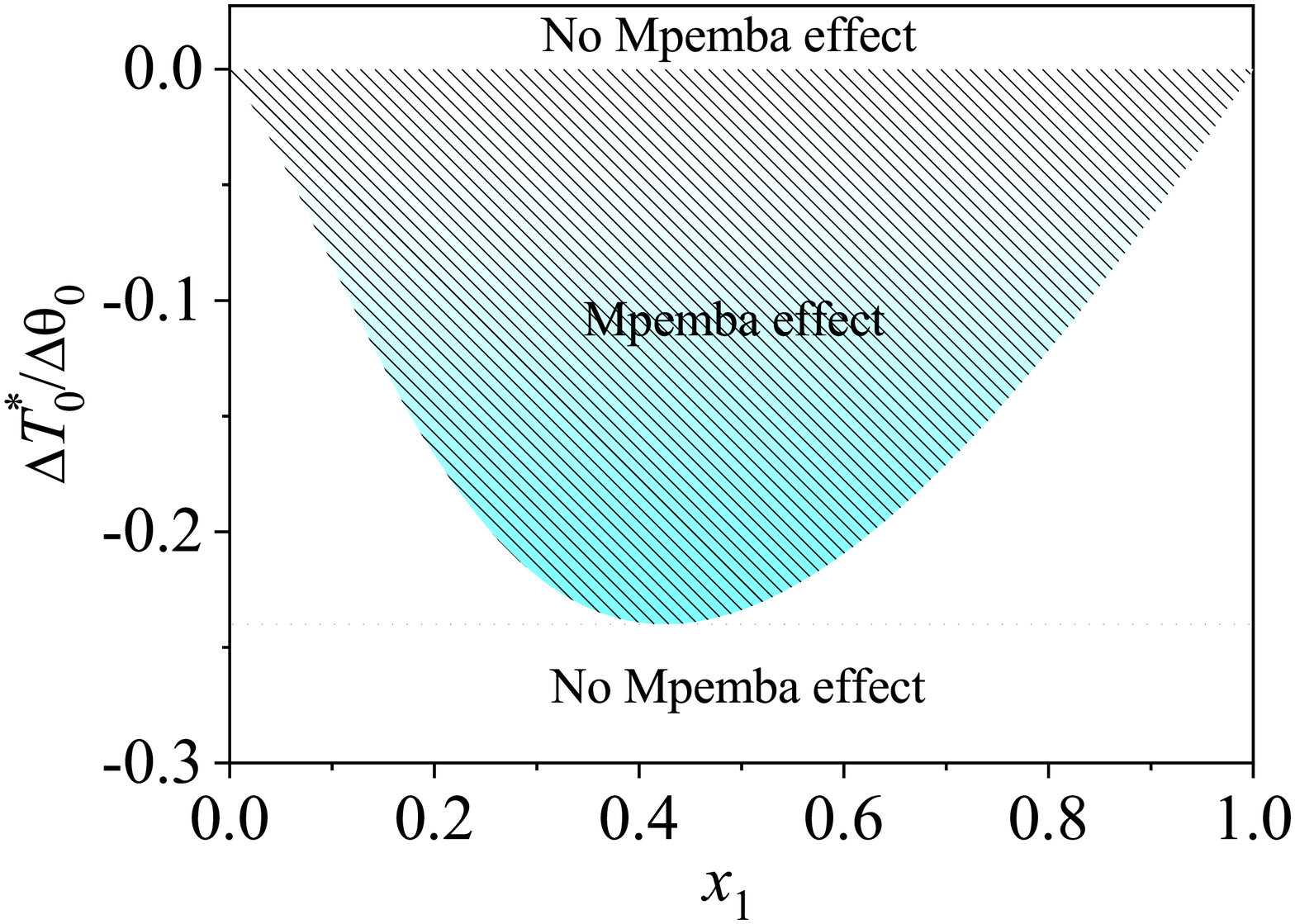}
	\includegraphics[width=0.45\textwidth]{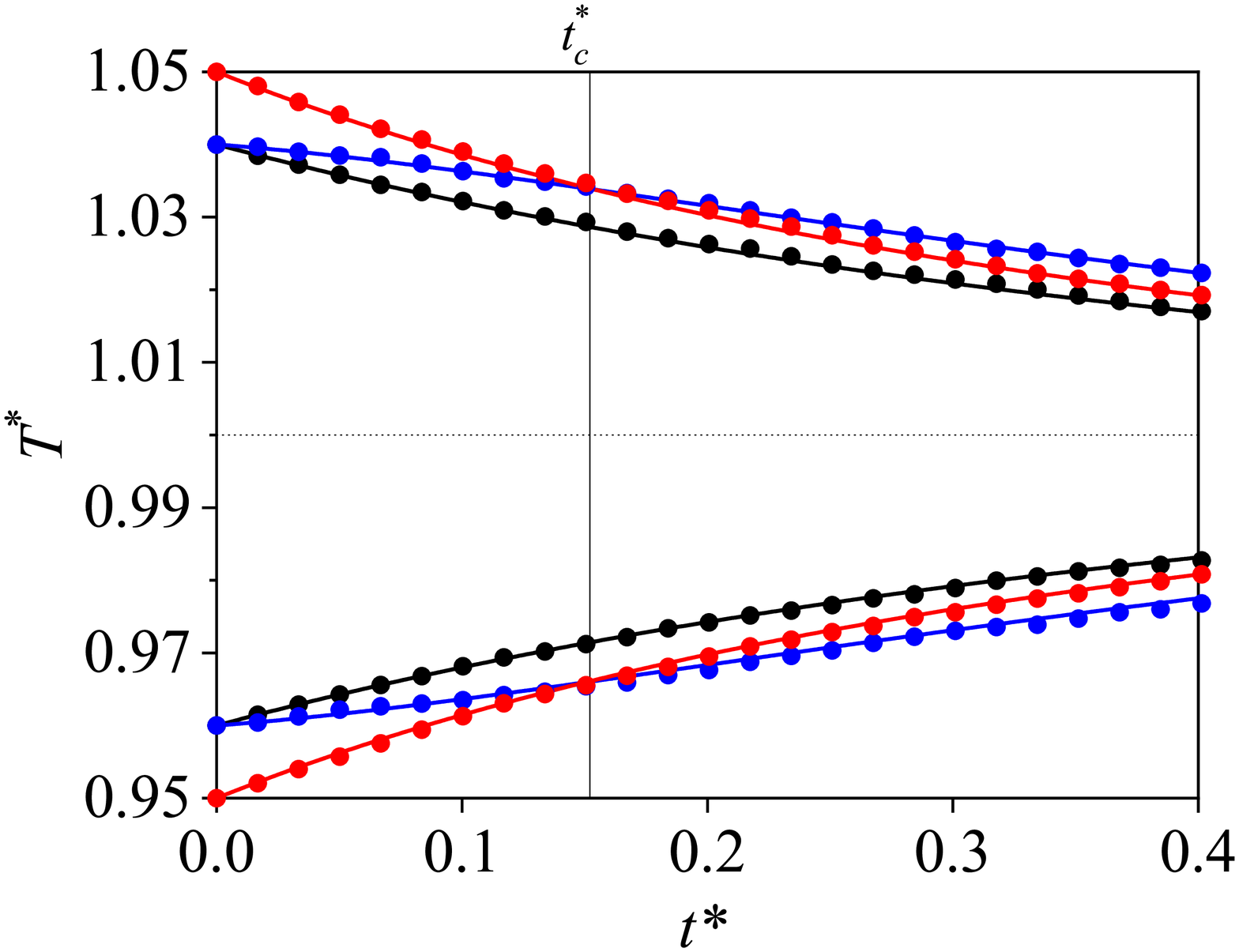}
	\caption{Upper panel: Phase diagram of the initial condition $\Delta T^*_0/\Delta\theta_0$ as a function of the concentration $x_1$. Lower panel: Relaxation of the (reduced) temperature $T^*$ over the time $t^*$ for $x_1=0.4$. The upper and lower curves correspond to the cooling and heating cases, respectively. Solid lines represents theoretical values and symbols DSMC data. The initial values of the control parameter $\Delta T^*_0/\Delta\theta_0\equiv (T^*_{A,0}-T^*_{B,0})/(\theta_{A,0}-\theta_{B,0})$ are $-0.1$ (A: red lines and symbols; B: blue lines and symbols), and $-0.5$ (A: red lines and symbols; B: black lines and symbols). The theoretical value of $t^*_c$ is also plotted with a vertical line. The remaining parameters in both panels are $d=3$,  $T_\text{ex}^*=1$, $m_1/m_2=5$, $\sigma_1/\sigma_2=1$, and $\phi=0.1$.}
	\label{fig3}
\end{figure}

In what follows, we will assume that the initial temperature of the state A is greater than that of the state B ($T_{A,0}^*>T_{B,0}^*$). From Eq. \eqref{18} we can now compute the possible crossing time $t^*_c$ of both trajectories. Given that, in the linear case, this time is invariant under the heating or cooling problem, both cases may be considered simultaneously. From the condition $\delta T^*_A(t^*_c)=\delta T^*_B(t^*_c)$, we obtain the expression
\beq
\label{19}
t^*_c=\frac{1}{\lambda_{-}-\lambda_{+}}\ln \frac{\mathcal L_{12}+(\mathcal L_{11}-\lambda_{-})\Delta T^*_0/\Delta \theta_0}{\mathcal L_{12}-(\lambda_+-\mathcal L_{11})\Delta T^*_0/\Delta \theta_0},
\eeq
where $\Delta T^*_0=T^*_{A,0}-T^*_{B,0}$ and $\Delta\theta_0=\theta_{A,0}-\theta_{B,0}$. In the linear theory, for given values of the parameters of the mixture, the crossover time $t^*_c$ depends on the initial conditions \emph{only} through the single control parameter $\Delta T^*_0/\Delta \theta_0$. Moreover, since $t^*_c\in \mathbb{R}^+$ and being aware of the inequality $\lambda_+>\lambda_-$, the argument of the logarithm in Eq.\ \eqref{19} shall fall within the interval $(0,1)$. Due to this restriction, the initial values must satisfy the following conditions
\beqa
\label{20}
\frac{\Delta T^*_0}{\Delta\theta_0}&\in&\left(0,\frac{\mathcal L_{12}}{\lambda_{-}-\mathcal L_{11}}\right)\quad \text{if} \quad \mathcal L_{12}<0\Leftrightarrow\gamma_1^*>\gamma^*_2,\nonumber\\
\frac{\Delta T^*_0}{\Delta\theta_0}&\in&\left(\frac{\mathcal L_{12}}{\lambda_{-}-\mathcal L_{11}},0\right)\quad \text{if} \quad \mathcal L_{12}>0\Leftrightarrow \gamma_1^*<\gamma^*_2.\nonumber\\
\eeqa
According to Eq.\ \eqref{20}, when $\mathcal L_{12}<0$ or equivalently $\gamma_1^*>\gamma_2^*$ ($\mathcal L_{12}>0$ or equivalently $\gamma_1^*<\gamma_2^*$), since $\lambda_{-}-\mathcal L_{11}<0$, the control parameter is $\Delta T^*_0/\Delta \theta_0> 0$ ($\Delta T^*_0/\Delta \theta_0< 0$) and its maximum (minimum) value for which the Mpemba effect can be observed is $\mathcal L_{12}/(\lambda_{-}-\mathcal L_{11})$. This quantity provides the phase diagram for the occurrence of the Mpemba effect, as shown in the upper panels of Figs.\ \ref{fig1}--\ref{fig3}. 
Here, it is important to study the singularity in the control parameter that emerge if $\theta_{A,0}=\theta_{B,0}$. In this case, the kinetic variables (partial temperatures) are not present in the early relative evolution of the macroscopic fields, $T_A$ and $T_B$, and therefore the Mpemba-like effect does not occur.

To illustrate the dependence of the required initial conditions on the parameters of the system, we consider a three-dimensional ($d=3$) system. In this case, a good approximation for the pair correlation functions $\chi_{ij}$ are given by~\cite{Boublik1970,Grundke1972,Lee1973}
\beq
\label{20.1}
\chi_{ij}=\frac{1}{1-\phi}+\frac{3}{2}\frac{\phi}{(1-\phi)^2}\frac{\sigma_i\sigma_jM_2}{\sigma_{ij}M_3}+\frac{1}{2}
\frac{\phi^2}{(1-\phi)^3}\left(\frac{\sigma_i\sigma_jM_2}{\sigma_{ij}M_3}\right)^2,
\eeq
where $M_\ell=\sum_i x_i\sigma_i^\ell$.

\begin{figure}[h]
	\centering
	\includegraphics[width=0.5\textwidth]{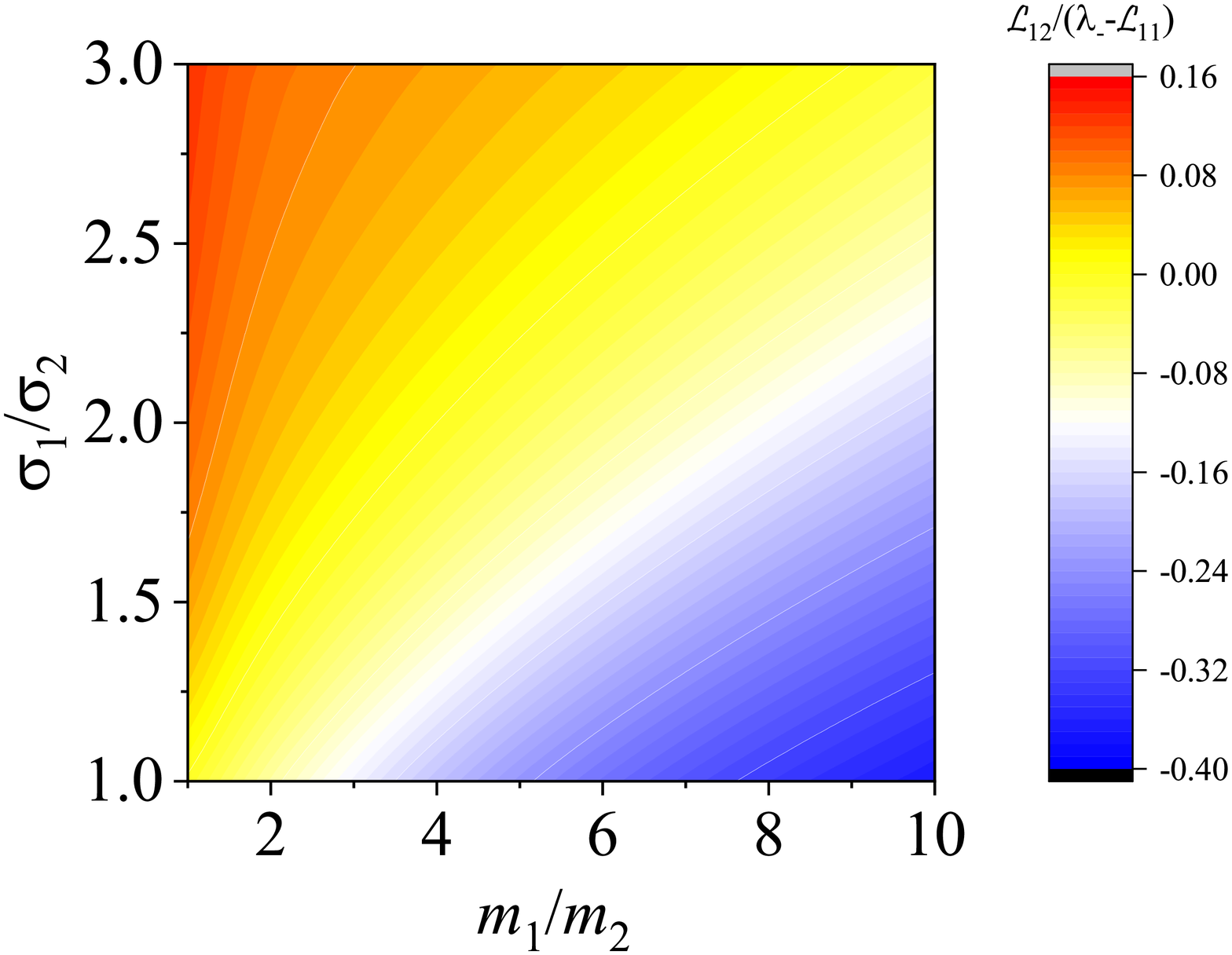}
	\caption{Density plot of the critical value $\mathcal L_{12}/(\lambda_{-}-\mathcal L_{11})$ as a function of the mass $m_1/m_2$ and size $\sigma_1/\sigma_2$ ratios for an equimolar mixture ($x_1=\frac{1}{2}$) of hard spheres ($d=3$). The remaining parameters are $T^*_\text{ex}=1$ and $\phi=0.1$.}
	\label{fig4}
\end{figure}

As said before, the upper panels of Figs.\ \ref{fig1}--\ref{fig3} show the phase diagram of the initial conditions $\Delta T^*_0/\Delta\theta_0$ as a function of the mass $m_1/m_2$ and size $\sigma_1/\sigma_2$ ratios and concentration $x_1$, respectively. We consider a binary molecular mixture of moderate density ($\phi=0.1$). If we focus on Fig.\ \ref{fig1}, we see that $\Delta T^*_0/\Delta\theta_0<0$ when $m_1>m_2$. For a better understanding, let us consider an equimolar binary mixture ($x_1=\frac{1}{2}$) with two components of identical diameters ($\sigma_1=\sigma_2$) but different masses ($m_1\neq m_2$). In these conditions, according to Eq.\ \eqref{2.1}, $\gamma_i^*\propto m_i^{-1}$ and so, $\gamma_1^*<\gamma_2^*$ when $m_1>m_2$. This means that the lighter component interchanges energy with the bath in a more efficient way than the heavier component. In addition, if we want a reduction in the relaxation time, the component whose partial temperature is further from $T_\text{ex}$ must be the one whose interaction with the bath is more effective. Since $\Delta T^*_0>0$ ($T_{A}^*>T_{B}^*$), then  $\Delta\theta_0<0$, namely, the initially hotter system has its kinetic energy more concentrated in the lighter component than in the heavier one. This conclusion agrees with the second condition of Eq.\ \eqref{20}.  As expected, the boundary between both regions, given by the extreme value $\mathcal L_{12}/(\lambda_{-}-\mathcal L_{11})$, decreases with increasing the mass ratio and hence, the discrimination between both species in the exchange of energy with the bath. A similar behavior is found in the upper panel of Fig.\ \ref{fig2}. Here, we vary the diameters of particles while keeping $m_1=m_2$. According to Eq.\ \eqref{2.1}, $\gamma_i^*\propto \sigma_i$ and so, $\gamma_1^*>\gamma_2^*$ when $\sigma_1>\sigma_2$. This implies that $\Delta T^*_0/\Delta\theta_0>0$ in agreement with the first condition of Eq.\ \eqref{20}. Conversely, the shape of the phase diagram shown in the upper panel of Fig.\ \ref{fig3} cannot be qualitatively explained with arguments based on individual properties (such as mass or size) but on collective behavior. As can be noted, the Mpemba effect manifests clearer when there are more particles that interact in a more efficient way with the bath. However, the mixture must also be diversified so there can be more discrepancy between both partial and total temperatures. An example of the competition between both mechanisms is plotted in the phase diagram of Fig.\ \ref{fig3} for $m_1=5m_2$.

The lower panels of Figs.\ \ref{fig1}--\ref{fig3} display the relaxation curves of the reduced temperature $T^*$ as a function of the scaled time $t^*$ for some of the mixture parameters considered in the phase diagrams described before. Three different initial conditions are chosen in every figure. Specific details of the initial conditions used in the above panels can be found in Table\ \ref{table1}. According to these initial values, the control parameter $\Delta T^*_0/\Delta\theta_0$ is greater than or less than 0, and within or without the region limited by $\mathcal L_{12}/(\lambda_{-}-\mathcal L_{11})$. In this way, the fulfillment of restrictions \eqref{20} is checked in both cooling and heating (inverse Mpemba effect) situations.  The solid lines are the theoretical results displayed in Eq.\ \eqref{18} and symbols refer to the results obtained via DSMC simulations.  We found an excellent agreement between theory and simulations in all the three cases, ensuring the accuracy of the Maxwellian approximation \eqref{9} to capture the trends of the Mpemba effect. Furthermore, the theoretical prediction for $t_c^*$ exhibits also an excellent agreement with simulations.

DSMC simulations has been carried out following similar steps as those carried out in Ref.~\onlinecite{Montanero1997}. At the initial state,
one assigns velocities to the particles drawn from Gaussian distributions at the desired partial temperatures. Since the system is assumed to be spatially homogeneous, the velocities of the particles
change only due to binary collisions. It includes two physical events: (i) collisions among particles and (ii) collisions of the particles with an external energy source (bath). In the case of event (i), we consider the same algorithm as proposed by Bird \cite{Bird1994} but, in this case, the collision rate is enhanced by a factor that accounts for the spatial correlations. \cite{Montanero2002} In the case of event (ii), we impose a simultaneous change of all velocities of particles every time step $\Delta t$. For a particle of species $i$ and velocity $\mathbf v$ the collision with the bath is given by
\begin{equation}
\label{20.1.1}
\mathbf v\to e^{-\gamma_i\Delta t}\mathbf v  +\left(\frac{6\gamma_iT_{\mathrm{ex}}}{m_i}\Delta t\right)^{\frac12}\mathbf w,
\end{equation}
where $\mathbf w$ is a random vector uniformly distributed in $[-1,1]^3$. In  Ref.~\onlinecite{Khalil2014}
it was shown that these two events (i) and (ii) give rise to the Boltzmann kinetic equation if $\Delta t$ is taken to be much smaller than the mean free time $\tau$ of inter-particle collisions. In our case we always take $\Delta t/\tau <10^{-3}$.

\begin{table}[h]
	\begin{tabular}{|c|c|c|c|c|c|c|}
		\hline
		& \multicolumn{2}{c|}{Figure 1} & \multicolumn{2}{c|}{Figure 2} & \multicolumn{2}{c|}{Figure 3} \\ \hline
		Color of lines and symbols & $T^*_0$      & $\theta_0$     & $T^*_0$      & $\theta_0$     & $T^*_0$      & $\theta_0$     \\ \hline
		\multicolumn{7}{|c|}{Cooling cases}                                                                                        \\ \hline
		Red                        & 1.05         & 1.06           & 1.05         & 1.11           & 1.05         & 1.01           \\ \hline
		Blue                       & 1.04         & 1.11           & 1.04         & 1.01           & 1.04         & 1.11           \\ \hline
		Black                      & 1.04         & 1.01           & 1.04         & 1.16           & 1.04         & 1.03           \\ \hline
		\multicolumn{7}{|c|}{Heating cases}                                                                                        \\ \hline
		Red                        & 0.95         & 0.94           & 0.95         & 0.89           & 0.95         & 0.99           \\ \hline
		Blue                       & 0.96         & 0.89           & 0.96         & 0.99           & 0.96         & 0.89           \\ \hline
		Black                      & 0.96         & 0.99           & 0.96         & 0.84           & 0.96         & 0.97           \\ \hline
	\end{tabular}
	\caption{Initial values of the (reduced) temperatures $T^*_0$ and temperature ratios $\theta_0$ used to generate the relaxation curves shown in the lower panels of Figs.\ \ref{fig1}--\ref{fig3}.}
	\label{table1}
\end{table}

As a complement of Figs.\ \ref{fig1}--\ref{fig3}, a density plot of the critical value  $\mathcal L_{12}/(\lambda_{-}-\mathcal L_{11})$ as a function of the mass and size ratios is plotted in Fig.\ \ref{fig4} for an equimolar mixture ($x_1=\frac{1}{2}$). Although these parameters have similar but opposite influences on the onset of the Mpemba effect, the graphic reveals that a discrimination in the diameters of particles (seen as a difference in the surface areas) has a more prominent role in the emergence of the phenomenon than in the masses (seen as a distinction in the inertial forces).

\section{Large Mpemba-like effect in molecular mixtures}
\label{sec4}

In the previous Section we have dealt with states which have been initially prepared in conditions close to thermal equilibrium. This has permitted us to linearize Eqs.\ \eqref{11} around the equilibrium solution and provide precise analytical results both for the time evolution of the temperature and for the crossing time. Here, we consider more general conditions, allowing the system to start away from equilibrium. In these cases we see that the relaxation curves may cross each other in a similar way to that described in Sec. \ref{sec2}. Unfortunately, no simple analytical expression for the crossing time $t^*_c$ is found and a more qualitative analysis is required to establish a necessary (but not sufficient) condition for the occurrence of the Mpemba effect.

In this Section we analyze crossovers in the temperature transitions from initial situations far away from equilibrium. Thus, the distances between the initial temperatures are assumed to be of the same order of the temperatures themselves. A remarkable fact of this kind of transitions is the asymmetry between the cooling and heating processes produced by the term $\xi_2-\xi_1$ of Eq.\ \eqref{7}. Given two initial temperatures $T_A>T_B$, both at the same distance from equilibrium ($T_A>T_\text{ex}>T_B;\; |T_A-T_\text{ex}|=|T_B-T_\text{ex}|$), one may think that the time to relax is exactly the same in both cases when the temperature ratios $\theta_{A,B}$ are also equally separated from their equilibrium values, in accordance with the linear theory of Sec.~\ref{sec3}. Nevertheless, on average, particles of system A move more energetically than those of system B so, the mean free time among collisions of species 1 and 2 of system A is shorter. Hence, the flux of linear momentum is more effective and, as a consequence, relaxation towards the external temperature turns out to be faster. This symmetry breaking results in a discrimination between cooling and heating processes in such a way that, for the same initial ratio $\Delta T^*_0/\Delta\theta_0$, Mpemba effect could only be observed in one of these scenarios.

Let us consider again two homogeneous states A and B arbitrarily far away from equilibrium. They are characterized by the  initial temperatures $T^*_{A,0}$ and $T_{B,0}^*$ and the temperature ratios $\theta_{A,0}$ and $\theta_{B,0}$. In what follows, for the sake of simplicity, we will suppose that the state A is initially hotter than B ($T^*_{A,0}>T^*_{B,0}$). Under this condition, a physically intuitive necessary condition for a crossover (Mpemba-like effect) in the relaxation curves of both temperatures is that the initially hotter system cools faster than the cooler one. This crossover is expected to happen in the early stage of the evolution where the system still puts away memory of its initial preparation. Following the arguments of Torrente \emph{et al.}~\cite{Torrente2019}, for short enough times, we can assume that the system is exponentially cooling with a characteristic rate roughly equal to the initial value of $-\Phi$ [see Eq.\ \eqref{11}]. Thus, a necessary condition for the presence of the Mpemba effect is
$\Phi(T_{B,0}^*,\theta_{B,0})>\Phi(T_{A,0}^*,\theta_{A,0})$. So, it seems that the function $\Phi(T^*,\theta)$, through its dependence on the variables $T^*$ and $\theta$, is the key quantity for determining when the Mpemba effect can occur.

Let us then analyze Eq.\ \eqref{12} to establish some restrictions to the initial conditions of the states A and B. The function $\Phi(T^*,\theta)$ is the sum of two functions $\Phi_1(T^*)$ and $\Phi_2(\theta)$. Thus, all the information about the relative behavior of $T_I^*$ ($I=$A,B) at the initial stages (for fixed $T_I^*$) falls on the function $\Phi_2(\theta_I)$. Next step is to ensure the functions $\Phi(T^*,\theta)$ behave monotonically with $\theta$, so that, we can establish a criterion for what the temperatures $T_A^*$ and $T_B^*$ will get closer or away from each other. Only the first option will be considered here as a simple way to attain Mpemba effect (in fact, there are other more complex ways the relaxation curves may cross as occurs for instance in the non-monotonic Kovacs-like relaxation.~\cite{Kovacs1979})

Therefore, to check the occurrence of the Mpemba effect we perform the derivative of $\Phi_2$ with respect to $\theta$ at fixed $T^*$. The  result is
\beq
\label{21}
\frac{\partial}{\partial\theta}\Phi_2=\frac{2x_1x_2(\gamma^*_2-\gamma_1^*)}{(x_2+x_1\theta)^2},
\eeq
which is always a positive (negative) function if $\gamma^*_2>\gamma^*_1$ ($\gamma^*_2<\gamma^*_1$). Consequently, assuming that the temperature evolves monotonically towards equilibrium, the presence of the Mpemba effect requires that the initial values satisfy the conditions
\beqa
\label{22}
\frac{\Delta T^*_0}{\Delta\theta_0}&>&0, \quad  \gamma_1^*>\gamma^*_2,\nonumber\\
\frac{\Delta T^*_0}{\Delta\theta_0}&<&0, \quad  \gamma_1^*<\gamma^*_2.
\eeqa
Equation \eqref{22} is in agreement with the results \eqref{20} derived for initial situations near to equilibrium. However, Eq.\ \eqref{22} does not constraint the regions of initial conditions that turn out in a crossover of temperatures. Namely, the difference between $\Phi(T_{A,0}^*,\theta_{A,0})$ and $\Phi(T_{B,0}^*,\theta_{B,0})$ must be properly chosen to be large enough.

According to Eq.\ \eqref{11}, the slope of the curve $T^*(t^*)$ is really the product $T^*\Phi$. Thus, one is tempted to conclude that the necessary condition for the presence of the Mpemba-like effect is $\Phi(T_{B,0}^*,\theta_{B,0})T_{B,0}>\Phi(T_{A,0}^*,\theta_{A,0})T_{A,0}$. On the other hand, since $T_{A,0}>T_{B,0}$, the Mpemba effect cannot occur \emph{unless} $\Phi(T_{B,0}^*,\theta_{B,0})>\Phi(T_{A,0}^*,\theta_{A,0})$. Therefore, the fulfillment of Eq.\ \eqref{22} is required to observe the Mpemba-like effect.

\begin{figure}[h]
	\centering
	\includegraphics[width=0.45\textwidth]{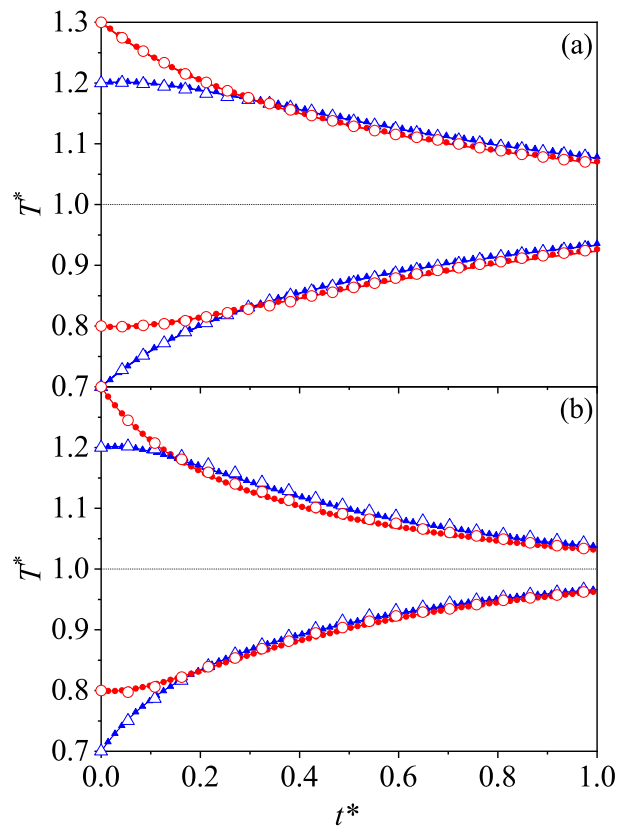}
	\caption{Evolution of the (reduced) temperature $T^*$ over the time $t^*$ for  $m_1/m_2=10$, $\sigma_1/\sigma_2=1$, $x_1=0.5$, $d=3$, and  $T_\text{ex}^*=1$. Solid lines represent theoretical values, filled symbols DSMC data, and open symbols MD data. The initial values of the control parameter $\Delta T^*_0/\Delta\theta_0\equiv (T^*_{A,0}-T^*_{B,0})/(\theta_{A,0}-\theta_{B,0})$ (A: red line and symbols; B: blue line and symbols) are $-0.2$ (cooling cases), $-0.25$ [heating case of panel (a)], and $-1/3$ [heating case of panel (b)]. Panel (a) corresponds to $\phi=0.00785$ and panel (b) to $\phi=0.1$.}
	\label{fig5}
\end{figure}

The large Mpemba-like effect for heating and cooling processes is plotted in Fig.\ \ref{fig5} for different initial conditions. Theoretical results are compared against DSMC and MD simulations. The MD simulations have been conducted as follows. The system is initially prepared in a spatially homogeneous state with each of the components of the mixture having Gaussian velocity distributions with different temperatures. The system then evolves using an event-driven algorithm. As in the case of DSMC simulations, two physical events are considered: (i) collisions among particles and (ii) collisions of the particles with the external bath. For the event (i), we proceeded as usual; see for instance Refs.~\onlinecite{allen2017computer,lu91}. In the event (ii), we impose a simultaneous change of all velocities of particles every time step $\Delta t$. This latter procedure is detailed in Eq.\ \eqref{20.1.1}.

In Fig.\ \ref{fig5}, we consider an equimolar ($x_1=\frac{1}{2}$) binary mixture of hard spheres ($d=3$) of the same size ($\sigma_1=\sigma_2$) and different masses ($m_1=10m_2$) for two different densities: $\phi=0.00785$ (very dilute system) and $\phi=0.1$ (moderately dense system). Lines are the theoretical results as derived from the Enskog equation, filled symbols refer to the results obtained via DSMC simulations, and open symbols to those obtained by means of MD simulations. When $\phi=0.00785$,  $\gamma_1^*=0.241$ and $\gamma^*_2=2.411$, while the friction parameters are $\gamma_1^*=0.445$ and $\gamma^*_2=4.451$ when $\phi=0.1$. Therefore, since $\gamma_1^*<\gamma_2^*$ in both cases, initial conditions must be chosen in such a way that $\Delta T^*_0/\Delta\theta_0<0$ (see Table\ \ref{table2} for more details). In addition, the functions $\Phi(T^*_I,\theta_I)$ are separately selected for the cooling and heating cases to enable the intersection of the respective temperature curves. It is quite apparent from the plots of Fig.\ \ref{fig5} that the Mpemba-like effect emerges in both (cooling and heating) relaxation problems, even when the relative differences in the initial temperatures are around $10\%$. Moreover, the panels (a) and (b) of Fig.\ \ref{fig5} highlight an excellent agreement between the Enskog theory and both DSMC and MD simulations in both the low-density regime ($\phi=0.00785$) and for moderate densities ($\phi=0.1$). This good agreement ensures once again the reliability of the Maxwellian approximation \eqref{9} as well as the accuracy of the molecular chaos hypothesis for studying this kind of relaxation process. The excellent agreement found in the crossing time $t_c^*$ and in the complete relaxation towards the final equilibrium state makes the Enskog kinetic theory a very reliable theory for modeling molecular fluids at moderate densities.

\begin{table}[h]
	\begin{tabular}{|c|c|c|c|c|}
		\hline
		& \multicolumn{2}{c|}{Panel (a)} & \multicolumn{2}{c|}{Panel (b)} \\ \hline
		Color of lines and symbols & $T^*_0$      & $\theta_0$      & $T^*_0$      & $\theta_0$      \\ \hline
		\multicolumn{5}{|c|}{Cooling cases}                                                          \\ \hline
		Red                        & 1.3          & 1.1             & 1.3          & 1.1             \\ \hline
		Blue                       & 1.2          & 1.6             & 1.2          & 1.6             \\ \hline
		\multicolumn{5}{|c|}{Heating cases}                                                          \\ \hline
		Red                        & 0.8          & 0.5             & 0.8          & 0.5             \\ \hline
		Blue                       & 0.7          & 0.9             & 0.7          & 0.8             \\ \hline
	\end{tabular}
	\caption{Initial values of the (reduced) temperatures $T^*_0$ and temperature ratios $\theta_0$ used to generate the relaxation curves shown in Fig.\ \ref{fig5}.}
	\label{table2}
\end{table}

\section{Mpemba-like effect in granular binary mixtures. Preliminary results}
\label{sec5}

\begin{figure}[h]
	\centering
	\includegraphics[width=0.45\textwidth]{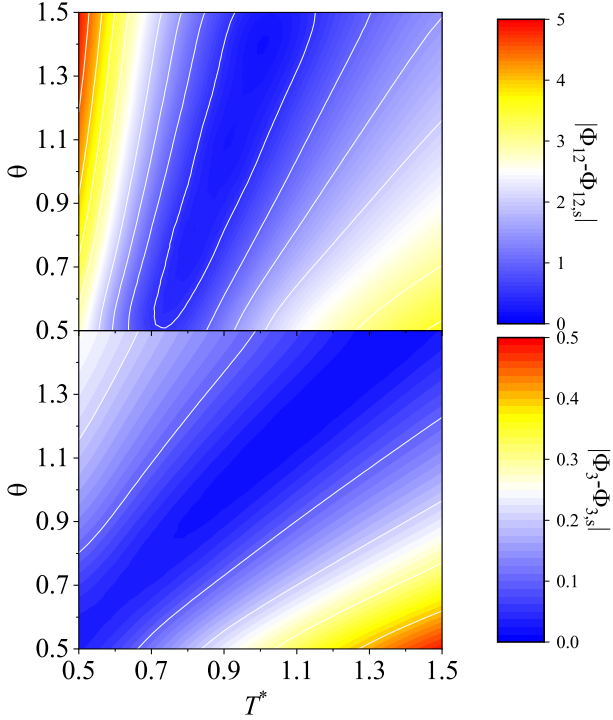}
	\caption{Density plot of the difference $\left|\Phi_{12}-\Phi_{12,\text{s}}\right|$ (top panel) and $\left|\Phi_3-\Phi_{3,\text{s}}\right|$ (bottom panel) as a function of the (reduced) temperature $T^*$ and temperature ratio $\theta$ for a granular mixture with a common coefficient of restitution $\alpha_{11}=\alpha_{22}=\alpha_{12}\equiv\alpha=0.9$. The parameters of the mixture are given by $m_1/m_2=10$, $\sigma_1/\sigma_2=1$, $x_1=0.5$, $d=3$, $\phi=0.1$, and  $T_\text{ex}^*=1$. Here, $\Phi_{12}=\Phi_1+\Phi_2$.}
	\label{fig6}
\end{figure}

	\begin{figure}[h!]
		\centering
		\includegraphics[width=0.45\textwidth]{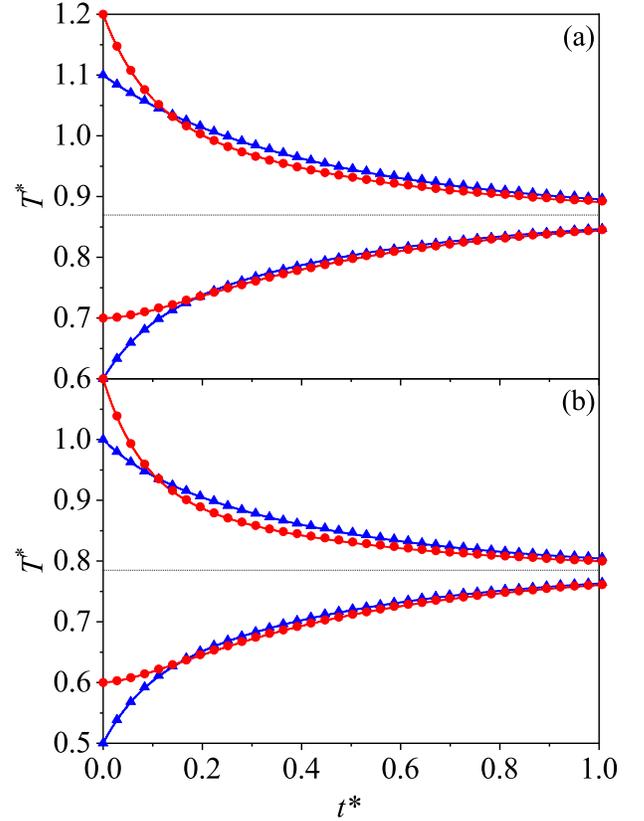}
		\caption{Evolution of the (reduced) temperature $T^*$ over the time $t^*$ for a granular mixture with a common coefficient of restitution $\alpha_{11}=\alpha_{22}=\alpha_{12}\equiv\alpha$. The parameters of the mixture are given by $m_1/m_2=10$, $\sigma_1/\sigma_2=1$, $x_1=0.5$, $d=3$, $\phi=0.1$, and  $T_\text{ex}^*=1$. Solid lines represent theoretical values and symbols DSMC data. The initial values of the control parameter $\Delta T^*_0/\Delta\theta_0\equiv (T^*_{A,0}-T^*_{B,0})/(\theta_{A,0}-\theta_{B,0})$ (A: red line and symbols; B: blue line and symbols) are $-0.25$ (cooling cases), $-1/3$ (heating cases). Panel (a) corresponds to $\alpha=0.9$ and panel (b) to $\alpha=0.8$.}
		\label{fig7}
	\end{figure}

We assume now that the components of the mixture have macroscopic dimensions (typically of the order of micrometers or larger), and so their collisions are \emph{inelastic}. We also assume that these particles (or grains) are in rapid flow conditions so that, they behave like a gas of activated collisional grains (granular gas).~\cite{BP04,Garzo2019} It is well-known that in this regime kinetic theory tools are appropriate to describe the dynamics of the system.

For granular mixtures suspended in a background fluid, the Enskog--Fokker--Planck equation \eqref{2} still applies, except that the Boltzmann--Enskog collision operator reads~\cite{Garzo2019}
\beqa
\label{5.1}
& & J_{ij}^{(\al_{ij})}[\mathbf{v}_1|f_i,f_j]=\chi_{ij} \sigma_{ij}^{d-1}\int d\mathbf{v}_{2}\int d\widehat{\boldsymbol {\sigma
}}\,\Theta (\widehat{{\boldsymbol {\sigma }}} \cdot {\mathbf
g}_{12})  \nonumber \\
& & \times (\widehat{\boldsymbol {\sigma }}\cdot {\mathbf g}_{12})
\left[\al_{ij}^{-2}f_{i}({\mathbf v}_{1}'')f_{j}({\mathbf v}_{2}'')-
f_{i}({\mathbf v}_{1})f_{j}({\mathbf v}_{2})\right],
\eeqa
where $\al_{ij}$ is the coefficient of normal restitution for collisions between particles of components $i$ and $j$. Here, the coefficient $\al_{ij}$ is assumed to be a positive constant smaller than or equal to 1. The limit case $\al_{ij}=1$ corresponds to elastic collisions. In Eq.\ \eqref{5.1}, the double primes denote the pre-collisional velocities $(\mathbf{v}_1'',\mathbf{v}_2'')$ yielding the post-collisional velocities $(\mathbf{v}_1,\mathbf{v}_2)$. They satisfy the collision rule:
\beq
\label{5.2}
\mathbf{v}_{1}''=\mathbf{v}_{1}-\mu_{ji}\frac{1+\al_{ij}}{\al_{ij}}(\widehat{{\boldsymbol {\sigma }}} \cdot {\mathbf g}_{12})\widehat{\boldsymbol {\sigma}},
\eeq
\beq
\label{5.2.1}
\mathbf{v}_{2}''=\mathbf{v}_{2}+\mu_{ij}\frac{1+\al_{ij}}{\al_{ij}}(\widehat{{\boldsymbol {\sigma }}} \cdot {\mathbf g}_{12})\widehat{\boldsymbol {\sigma}}.
\eeq
The operator $J_{ij}^{(\al_{ij})}$ denotes the \emph{inelastic} Enskog-Boltzmann collision operator. When $\al_{ij}=1$, its elastic version $J_{ij}^{(1)}$ is  given by Eq.\ \eqref{1.1}.

The study of the Mpemba-like effect for driven granular mixtures follows similar mathematical steps as those made in the previous Sections for molecular mixtures. Thus, the set of differential equations \eqref{11} provides the evolution of the (reduced) temperature $T^*$ and the temperature ratio $\theta$. However, the final forms of the functions $\Phi$ and $\Psi$ for inelastic collisions [see Eqs.\ \eqref{a2}--\eqref{a7} of the Appendix \ref{appA}] appearing in the above set of differential equations are much  more intricate than those obtained for molecular gases.

Nonetheless, preliminary straightforward results can be derived if we realize that the dependence of the function $\Phi(T^*,\theta)=\Phi_1(T^*)+\Phi_2(\theta)+\Phi_3(T^*,\theta)$ on inelasticity is fully encoded in the cooling term $\Phi_3$. This cooling term vanishes for elastic collisions ($\Phi_3=0$ when $\al_{ij}=1$). Thus, to establish some criterion on the emergence of the Mpemba-like effect, the function $\Phi$ is conveniently separated into its entirely molecular part $\Phi_{12}\equiv \Phi_1+\Phi_2$ and the granular term $\Phi_3$.
	
Let us consider again two different homogeneous samples $A$ and $B$ at different initial granular temperatures $T^*_{I,0}$ and temperature ratios $\theta_{I,0}$, where $I=A,B$. In order to compare the relative behaviour between the two slopes $\Phi_A$ and $\Phi_B$ at the initial stages of the evolution, the steady values of the molecular $\Phi_1(T^*_\text{s})+\Phi_2(\theta_\text{s})$ and the granular $\Phi_3(T^*_\text{s},\theta_\text{s})$ terms are subtracted from their non-steady slopes $\Phi_1(T^*)+\Phi_2(\theta)$ and $\Phi_3(T^*,\theta)$, respectively. Thus, we can quantify the influence of the granular terms on the relative distance between the relaxation curves and, hence, on the onset of the Mpemba effect.

The set of coupled equations for obtaining the steady forms of both $T^*_\text{s}$ and $\theta_\text{s}$ are given by Eqs.\ \eqref{a9} and \eqref{a10}. It is quite apparent from Fig.\ \ref{fig6} that the granular term $\Phi_3$ has substantially less influence on the relative behaviour of the temperature relaxation of two given samples than the molecular counterpart $\Phi_1+\Phi_2$ at moderate values of the coefficients of restitution $\alpha_{ij}$. In this way, similar conditions to those previously obtained for driven molecular mixtures can be established for granular mixtures to chose the initial values of $T^*$ and $\theta$ for the occurrence of the Mpemba-like effect.

\begin{table}[h!]
		\begin{tabular}{|c|c|c|c|c|}
			\hline
			& \multicolumn{2}{c|}{Panel (a)} & \multicolumn{2}{c|}{Panel (b)} \\ \hline
			Color of lines and symbols & $T^*_0$      & $\theta_0$      & $T^*_0$      & $\theta_0$      \\ \hline
			\multicolumn{5}{|c|}{Cooling cases}                                                          \\ \hline
			Red                        & 1.2          & 0.9             & 1.1          & 0.8             \\ \hline
			Blue                       & 1.1          & 1.3             & 1.0          & 1.2            \\ \hline
			\multicolumn{5}{|c|}{Heating cases}                                                          \\ \hline
			Red                        & 0.7          & 0.5             & 0.6          & 0.4             \\ \hline
			Blue                       & 0.6          & 0.8             & 0.5          & 0.7             \\ \hline
		\end{tabular}
		\caption{Initial values of the (reduced) temperatures $T^*_0$ and temperature ratios $\theta_0$ used to generate the relaxation curves shown in Fig.\ \ref{fig7}.}
		\label{table3}
	\end{table}

Here, as an illustration of the Mpemba-like effect in granular mixtures, the time evolution of $T^*$ is plotted in Fig.\ \ref{fig7} for heating and cooling processes. For the sake of comparison, we consider a binary granular suspension with the same mechanical properties as those considered in panel (b) of Fig.\ \ref{fig5}, except that now the collisions are inelastic. Two different values of the (common) coefficient of restitution $\alpha\equiv \alpha_{11}=\alpha_{22}=\alpha_{12}$ are selected: (a) $\alpha=0.9$ and (b) $\alpha=0.8$. Lines are the theoretical results derived from the Enskog equation conveniently adapted to inelastic collisions (see the Appendix \ref{appA}) and symbols refer to the results obtained via DSMC simulations. Since the friction parameters are $\gamma_1^*=0.445$ and $\gamma_2^*=4.451$ in both cases, similar arguments than those derived in the molecular case are set out for the initial conditions to satisfy the relation $\Delta T^*_0/\Delta \theta_0<0$ (more details can be found in Table \ref{table3}). Figure \ref{fig7} illustrates the emergence of the Mpemba-like effect (and its inverse counterpart) in granular gases when the initial conditions are relatively far away from each other (large Mpemba-like effect). In addition, panels (a) and (b) of Fig.\ \ref{fig7} show an excellent agreement between the Enskog theory and DSMC simulations ensuring again the reliability of the Maxwellian approximation used to compute the partial production rates $\xi_i$ given in Eq.\ \eqref{a1}.

\section{Discussion}
\label{sec6}

In summary, we have observed a Mpemba-like effect in a molecular binary mixture in contact with a thermal reservoir. As usual,~\cite{RL77,K92,Koch2001,zw01} the bath acts on molecules as they were Brownian particles, i.e. the interaction between gas particles and the thermal reservoir (or background fluid) is accounted for by two forces: a (deterministic) viscous drag force proportional to the velocity of the particles and a stochastic force. Moreover, based on numerical and experimental results carried out in the gas-solid-flows literature, \cite{Yin2009,Yin2009a,Holloway2009} the friction coefficients $\gamma_i$ ($i=1,2$) have been chosen to distinguish between components of the mixture through their dependence on the mechanical properties of particles (masses $m_i$ and diameters $\sigma_i$) and on the partial $\phi_i$ and global $\phi=\phi_1+\phi_2$ volume fractions. This discrimination couples the evolution of the total temperature $T(t)$ with the ratio  of partial temperatures $\theta(t)=T_1(t)/T_2(t)$ giving rise to the emergence of memory effects. Namely, the time evolution of $T(t)$ is not autonomous but is coupled to $\theta(t)$. One of the most popular problems in which memory effects are notorious is the so-called Mpemba effect,~\cite{Mpemba1969} namely, when an initially hotter (cooler) system cools (heats) sooner.

To observe this effect, two identical samples $A$ and $B$ (namely, sharing the same values of masses, diameters, composition, and volume fraction) are initially prepared in isotropic Maxwellian velocity distribution functions at different temperatures ($T_{A,0}$ and $T_{B,0}$) and temperature ratios ($\theta_{A,0}$ and $\theta_{B,0}$). These samples are in contact with a thermal reservoir at temperature $T_\text{ex}$. Starting from the above initial conditions, we let the samples evolve until they reach the equilibrium state where energy equipartition holds: $T_i=T_{1,i}=T_{2,i}=T_\text{ex}$ ($i=A,B$). During this transient period, particles of the mixture collide among themselves and with the bath exchanging energy in different ways for each component. If we suitable chose the initial values of the total and the partial temperatures, the curves associated with the relaxation of the temperatures $T_A(t)$ and $T_B(t)$ may cross at a given time  $t_c$ before reaching the equilibrium state (the so-called crossover time). Contrary to other works on this topic, no cumulants~\cite{Lasanta2017} (measuring the deviations of the distribution functions from their Maxwellian forms) nor the inclusion of a nonlinear drag force~\cite{Santos2020} are needed to explain the Mpemba effect and hence, the magnitude of the effect may be increased.

The starting point of our theoretical approach has been the Enskog kinetic equation \eqref{1} in combination with the Fokker-Planck term accounting for the interaction between gas particles and the thermal reservoir. From this equation, the evolution equations for the total temperature $T(t)$ and the temperature ratio $\theta(t)$ have been derived. To get explicit results, the partial production rates $\xi_1$ and $\xi_2$ appearing in the evolution equation \eqref{7} of $\theta$ have been estimated by replacing the exact distribution functions $f_i(\mathbf{v};t)$ by their Maxwellian forms \eqref{9}.

The evolution equations \eqref{11} for the reduced quantities $T^*=T/T_{\text{ex}}$ and $\theta$ have been first analytically solved for situations close to equilibrium. This has allowed us to obtain \textit{explicit} expressions for the (reduced) crossing time $t^*_c$ and the critical value of the initial temperature difference [see Eqs.\ \eqref{19} and \eqref{20}]. In addition, the numerical solution of the set of equations \eqref{11} provides the dependence of $T^*(t^*)$ and $\theta(t^*)$ on the parameters of the mixture. An illustration of the above results is displayed in Figs.\ \ref{fig1}--\ref{fig3} where we have varied the mass $m_1/m_2$ and diameter $\sigma_1/\sigma_2$ ratios and the composition $x_1$, respectively. The comparison between those theoretical (approximate) predictions with the DSMC results shows an excellent agreement for the whole range of parameters studied.

As a complement of the previous study, we have analyzed the Mpemba effect when the initial states of the samples are far from equilibrium, the so-called large Mpemba effect. In this situation, no analytical solution is admitted and \emph{only} qualitative predictions can be achieved. For the crossover to happen, a necessary criterion for the sign of the initial fraction $\Delta T^*_0/\Delta\theta_0=(T^*_{A,0}-T^*_{B,0})/(\theta_{A,0}-\theta_{B,0})$ has been established. This criterion is based on the efficiency or rapidity (measured through the comparison of the two drag coefficients $\gamma^*_i$) of each one of the partial temperatures to reach equilibrium. Two examples of cooling and heating relaxation processes for a dilute and a moderately-dense system has been plotted in Fig.\ \ref{fig5}. In particular, the Mpemba-like effect has been shown to take place even when the relative initial temperature difference is around 10$\%$. Moreover, an excellent agreement between theoretical results and both DSMC and MD simulations has been also found.

Finally, we have also considered driven granular mixtures, namely, a collection of discrete macroscopic particles of different sizes. Due to their macroscopic dimensions, in contrast to molecular mixtures, the collisions between the different components of the mixture are inelastic. As expected, the Mpemba-like effect is also present when collisions in the binary mixture are inelastic. However, given that the forms of the functions $\Phi=\Phi_1+\Phi_2+\Phi_3$ and $\Psi$ appearing in the evolution equations obeying $T^*$ and $\theta$, respectively, are more complex than those derived for elastic collisions, it is not easy to find clean conditions for the occurrence of the Mpemba effect. On the other hand, since the impact of the granular new term $\Phi_3$ (which vanishes for molecular mixtures) on the relaxation of the temperature is smaller than that of the pure molecular contributions $\Phi_1+\Phi_2$ for not too strong inelasticities, one can conclude that the conditions for the occurrence of the Mpemba-like effect in granular mixtures are quite similar to those found for driven molecular mixtures. In any case, a more careful analysis is needed to confirm the above conclusion. We plan to carry out a more exhaustive study on the necessary conditions for the onset of the Mpemba-like effect in driven granular mixtures in the near future.

\acknowledgments

The work of R.G.G. and V.G. has been supported by the Spanish Government through Grant No. FIS2016-76359-P and by the Junta de Extremadura (Spain) Grant No. GR18079, partially financed by ``Fondo Europeo de Desarrollo Regional'' funds. The research of R.G.G. also has been supported by the predoctoral fellowship BES-2017-079725 from the Spanish Government.

\appendix

\section{Expressions for driven granular mixtures}
\label{appA}

\begin{widetext}

In this Appendix, we display the expressions of the functions $\Phi$ and $\Psi$ for driven granular mixtures, namely, when collisions between particles of the component $i$ and $j$ are inelastic. For smooth hard spheres, the inelasticity of collisions are characterized by the (constant) coefficients of restitution $\alpha_{ij}\leq 1$. In this case, the expressions of the partial production rates $\xi_i$ in the Maxwellian approximation \eqref{9} are given by~\cite{Garzo2019}
\beqa
\label{a1}
\xi_1&=&\frac{\sqrt{2}\pi^{\left(d-1\right)/2}}{d\Gamma\left(\frac{d}{2}\right)}n_1 \chi_{11} \sigma_{1}^{d-1}\left(\frac{2T_1}{m_1}\right)^{1/2}\left(1-\al_{11}^2\right)+\frac{4\pi^{\left(d-1\right)/2}}{d\Gamma\left(\frac{d}{2}\right)}n_2 \mu_{21}\chi_{12} \sigma_{12}^{d-1}\left(\frac{2T_1}{m_1}+\frac{2T_2}{m_2}\right)^{1/2}\nonumber\\
& & \times (1+\al_{12})\Bigg[1-\frac{\mu_{21}}{2}\left(1+\al_{12}\right)\Bigg(1+\frac{m_1T_2}{m_2T_1}\Bigg)\Bigg].
\eeqa
The expression for $\xi_2$ can be easily obtained from Eq.\ \eqref{a1} by making the change $1\leftrightarrow 2$. In dimensionless variables, the time evolution of $T^*$ and $\theta$ can be written in the form \eqref{11} where
\beqa
\label{a2}
\Phi(T^*,\theta)=\Phi_1(T^*)+\Phi_2(\theta)+\Phi_3(T^*,\theta), \quad \Psi(T^*,\theta)&=&\Psi_1+\Psi_2(T^*,\theta)+\Psi_3(T^*,\theta).
\eeqa
Here, we have introduced the quantities
\beq
\label{a3}
\Phi_1(T^*)=\frac{2}{T^*}\left(x_1\gamma_1^*+x_2 \gamma_2^*\right), \quad \Phi_2(\theta)=-2 \frac{x_1 \gamma_1^* \theta+x_2 \gamma_2^*}{1+x_1(\theta-1)}, \quad \Phi_3(T^*,\theta)=-\xi^*(T^*,\theta),
\eeq
\beq
\label{a4}
\Psi_1=-2(\gamma_1^*-\gamma_2^*), \quad \Psi_2(T^*,\theta)=2\left(\gamma^*_1-\gamma_2^*\theta\right)\frac{1+x_1(\theta-1)}{\theta T^*},
\quad \Psi_3(T^*,\theta)=\xi_2^*(T^*,\theta)-\xi_1^*(T^*,\theta),
\eeq
where
\beq
\label{a5}
\xi^*=\frac{x_1\theta\xi_1^*+x_2 \xi_2^*}{1+x_1(\theta-1)},
\eeq
\beqa
\label{a6}
\xi_1^*&=&\frac{\sqrt{2}\pi^{\left(d-1\right)/2}}{d\Gamma\left(\frac{d}{2}\right)}x_1 \chi_{11}\Big(\frac{\sigma_1}{\sigma_{12}}\Big)^{d-1}
\sqrt{\frac{\theta T^*}{2\mu_{12}\Big[1+x_1(\theta-1)\Big]}}\left(1-\al_{11}^2\right)+\frac{4\pi^{\left(d-1\right)/2}}{d\Gamma\left(\frac{d}{2}
\right)}x_2 \chi_{12}\nonumber\\
& & \times \sqrt{\frac{\mu_{21}}{\mu_{12}}}
\sqrt{\frac{T^*}{2}\frac{\mu_{12}+\mu_{21}\theta}{1+x_1(\theta-1)}}
\left(1+\al_{12}\right)\Big[1-\frac{1+\al_{12}}{2}\Big(\mu_{21}+\mu_{12}\theta^{-1}\Big)\Big],
\eeqa
\beqa
\label{a7}
\xi_2^*&=&\frac{\sqrt{2}\pi^{\left(d-1\right)/2}}{d\Gamma\left(\frac{d}{2}\right)}x_2 \chi_{22}\Big(\frac{\sigma_2}{\sigma_{12}}\Big)^{d-1}
\sqrt{\frac{T^*}{2\mu_{21}\Big[\theta+x_2(1-\theta)\Big]}}\left(1-\al_{22}^2\right)+\frac{4\pi^{\left(d-1\right)/2}}{d\Gamma\left(\frac{d}{2}
\right)}x_1 \chi_{12}\nonumber\\
& & \times \sqrt{\frac{\mu_{12}}{\mu_{21}}}
\sqrt{\frac{T^*}{2}\frac{\mu_{12}+\mu_{21}\theta}{1+x_1(\theta-1)}}
\left(1+\al_{12}\right)\Big[1-\frac{1+\al_{21}}{2}\Big(\mu_{12}+\mu_{21}\theta\Big)\Big].
\eeqa
For elastic collisions ($\al_{11}=\al_{22}=\al_{12}=1$), Eqs.\ \eqref{a4}--\eqref{a7} reduce to Eqs.\ \eqref{12}--\eqref{15} since $\Phi_3=0$ and
\beq
\label{a8}
\Psi_3=\xi_2^*-\xi_1^*=\frac{8\pi^{\left(d-1\right)/2}}{d\Gamma\left(\frac{d}{2}
\right)}\chi_{12} \sqrt{\frac{T^*}{2}\frac{\mu_{12}\mu_{21}\left(\mu_{12}+\mu_{21}\theta\right)}{1+x_1(\theta-1)}}\Big(x_1-x_1\theta-x_2+x_2\theta^{-1}\Big).
\eeq

In the long-time limit, the steady forms of both $T_s^*$ and $\theta_s^*$ can be obtained by solving the set of coupled equations
\beq
\label{a9}
\frac{2}{T_s^*}\left(x_1\gamma_1^*+x_2 \gamma_2^*\right)-2 \frac{x_1 \gamma_1^* \theta_s+x_2 \gamma_2^*}{1+x_1(\theta_s-1)}=\xi_s^*,
\eeq
\beq
\label{a10}
-2(\gamma_1^*-\gamma_2^*)+2\left(\gamma_1^*-\gamma_2^*\theta_s\right)\frac{1+x_1(\theta_s-1)}{\theta_s T_s^*}=\xi_{1s}^*-\xi_{2s}^*.
\eeq
For elastic collisions, the solution to Eqs.\ \eqref{a9} and \eqref{a10} is simply given by $T_{s}^*=\theta_{s}^*=1$ (energy equipartition). However, for inelastic collisions, energy equipartition does not hold and $T_s^*$ and $\theta_s^*$ have a complex dependence on the parameter space of the problem. An study on this dependence has been carried out in Ref.\ \onlinecite{Gonzalez2020} for a binary mixture and in Ref.\ \onlinecite{OBB20} for a multicomponent mixture.

\end{widetext}

\textbf{DATA AVAILABILITY}

The data that support the findings of this study are available from the corresponding author upon reasonable request.

%\bibliography{Mpemba}

%\end{document}

%

%\bibliography{Mpemba}

\end{document}